\documentclass[10pt]{article}

\usepackage{amsmath, amsfonts, amssymb, bm}

\usepackage[utf8]{inputenc}
\usepackage[T1]{fontenc}
\usepackage{lmodern}

\usepackage{hyperref}
\usepackage{graphicx, bm, amsmath, color}
\usepackage{bbold}
\usepackage{cite}
\usepackage{wasysym}
\usepackage{verbatim}
\usepackage{multicol}

\usepackage{epstopdf}
\usepackage{animate}
\usepackage{xmpmulti}
\usepackage{centernot}
\usepackage{multirow}
\usepackage{tikz}
\usepackage{graphicx}
\usepackage{float}
\usepackage{mathtools}
\usepackage{comment}
\usepackage{braket}

\makeatletter

\usepackage{amsmath} 
\newcommand{\be}{\begin{equation}}
\newcommand{\ee}{\end{equation}}
\newcommand{\beq}{\begin{eqnarray}}
\newcommand{\eeq}{\end{eqnarray}}
\newcommand{\ba}{\begin{align}}
\newcommand{\ea}{\end{align}}

\oddsidemargin=
-0.45cm
\evensidemargin=
-0.45cm
\begin{document}
  \begin{center}
    \baselineskip 24 pt {\LARGE \bf Thermodynamics of the $q$-deformed Kittel--Shore model}
  \end{center}

  \bigskip
  \bigskip
  \begin{center}
    {\sc Víctor Mariscal $^{1}$ and J. Javier Relancio$^{1,2}$}

    \medskip
    {$^{1}$Departamento de Matem\'aticas y Computaci\'on, Universidad de Burgos, 09001 Burgos, Spain}

    {$^{2}$Centro de Astropart\'{\i}culas y F\'{\i}sica de Altas Energ\'{\i}as (CAPA), Universidad de Zaragoza, Zaragoza 50009, Spain}
    \medskip

    e-mail:
    {\href{mailto:vmariscal@ubu.es}{vmariscal@ubu.es}, \href{mailto:jjrelancio@ubu.es}{jjrelancio@ubu.es}}
  \end{center}

  \medskip

  \begin{abstract}
    The Kittel--Shore Hamiltonian characterizes $N$ spins with identical long-range interactions, and the $\mathfrak{su}(2)$ coalgebra has been proven to be a symmetry  of this model, which can be exactly solved. By using quantum groups and, in particular, $\mathfrak{su}_{q}(2)$, this Hamiltonian was deformed. In this work, we study the thermodynamic properties of this deformed model for spin-$1/2$ particles. In particular, we discuss how this deformation affects the specific heat, magnetic susceptibility, magnetisation, and phase transitions as a function of the parameter $q$ of the deformation and compare them with those of the undeformed model. Deformation was found to shift the thermodynamic behaviours to higher temperatures and alter the phase transitions. {Additionally, we extend the framework to complex deformations on the unit circle, $q = e^{i\theta}$, for $q$ not being a root of unity. For an $N=4$ tetrahedral cluster, this phase establishes a thermodynamic equivalence between a second-neighbor spin-Hamiltonian and the $q$-deformed quantum model. At least, we have found experimental compounds well reproduced by the Kittel--Shore model for $N=4$. Validating our effective algebraic model against exact diagonalisation data for the $\text{Cu}_2\text{Te}_2\text{O}_5X_2$ ($X=\text{Cl}, \text{Br}$) compound family demonstrates that $\theta$ accurately reproduces the structural anisotropy.} 
  \end{abstract}

  \medskip
  \medskip
 


  \tableofcontents

  \section{Introduction}

  The Kittel--Shore (KS) model~\cite{kittel1965development} describes an infinite-range
  spin system in which $N$ spins interact identically. It can be regarded as a
  vectorial extension of the Ising model: while Ising spins only possess a $z$-component,
  KS spins span the full $\mathfrak{su}(2)$ algebra. Because of this relationship,
  the KS Hamiltonian has been used in several Ising-based analyses, including the
  demonstration of the exactness of the molecular field approximation~\cite{kittel1965development}
  and the study of first-order transitions in $J=1$ Ising systems with crystal-field
  splitting~\cite{capel1967possibility}. Finite-size Ising simulations have also
  exploited KS results to infer higher-dimensional behaviour~\cite{nagle1970numerical},
  and mappings of quantum spin partition functions to Ising models have been formulated
  through generalised Trotter product expansions~\cite{suzuki1976relationship,trotter1959product}.
  Further connections appear in studies of long-range quantum magnetism and
  quantum information~\cite{van1993heisenberg,czachor2002verification}.

  The exact solvability of the KS model follows from the decomposition of the
  tensor product of $N$ irreducible $U(\mathfrak{su}(2))$ representations into
  invariant subspaces~\cite{kittel1965development}. It serves as a canonical and
  analytically tractable representative of long-range interacting spin models and
  exhibits a ferromagnetic phase transition that develops slowly as the system size
  increases ~\cite{kittel1965development}. These properties support finite-size scaling
  analyses in infinitely coordinated systems~\cite{botet1982size,botet1983large},
  in accordance with Fisher’s and Barber’s scaling hypothesis~\cite{cardy1988current}.
  Arbitrary spin extensions were developed in~\cite{al1998exact,van1993heisenberg,czachor2002verification},
  and antiferromagnetic couplings were examined in~\cite{al1998exact,czachor2008energy}.

  It is worth emphasising that the KS Hamiltonian corresponds to the Heisenberg
  XXX model on the complete graph~\cite{bjornberg2020quantum}, that is, the long-range
  generalisation of the XXX model with a fully symmetric constant coupling and
  thus a standard mean-field Hamiltonian~\cite{fannes1980equilibrium} (once $I\!\rightarrow
  \! I/N$). Models on the complete graph exhibit strong analytical accessibility.
  In the spin-$1/2$ and spin-1 cases, integral formulas for the thermodynamic-limit
  partition function (ferromagnetic regime), explicit magnetisation, free energy,
  and critical exponents are available~\cite{van1993heisenberg,bjornberg2020quantum,ryan2023class,toth1990phase,penrose1991bose,bjornberg2016free,alon2021mean,bjornberg2023heisenberg}.
  In the spin-1 setting, KS-type Hamiltonians have also been studied as
  bilinear–biquadratic models~\cite{papanicolaou1986ground,jakab2018bilinear}. Rotational
  and permutational symmetries are fundamental in deriving most of these results.

  Geometrically, the KS Hamiltonian ($H_{KS}$) describes $N$ spins at the
  vertices of a $(N-1)$-dimensional simplex, becoming effectively infinite dimensional
  in the thermodynamic limit $N\to\infty$. {It is worth noting that for $N=2$ (dimer) and $N=3$ (equilateral triangle), the KS model is equivalent to the nearest-neighbor Heisenberg model, including periodic boundary conditions in the latter case. The distinct connectivity of the KS model, where each spin interacts with all others, becomes evident starting from $N=4$ (a tetrahedral geometry).}
  Small systems with $N=2,3,4$ have been used to model ultra-small magnetic clusters~\cite{ciftja1999equation}
  and appear in quantum-dot-based quantum computation proposals~\cite{loss1998quantum,woodworth2006few}.

  The classical KS model was analysed in terms of the dynamics and integrals of motion.
  Its time evolution was studied in~\cite{liu1991dynamics}, invariants and action-angle
  variables were obtained in~\cite{magyari1987integrable}, and analytical
  autocorrelation functions were derived in~\cite{muller1988high}. Anisotropic generalisations
  (equivalent-neighbour XYZ) and their correlation functions were addressed in~\cite{dekeyser1979time,lee1984time}.

  Quantum groups, such as $U_{q}(\mathfrak{su}(2))$, underlie the symmetries of integrable
  spin chains like the XXZ Heisenberg model {with {specific} open boundary conditions}
  for $N$ spin-$\frac{1}{2}$ particles, with the deformation parameter $q$ linked
  to the anisotropy $\delta$ by $\delta = (q + q^{-1})/2$ and an additional boundary
  term $(q - q^{-1})(\sigma_{z}^{1}- \sigma_{z}^{N})/2$~\cite{vladimir1986drinfeld,jimbo1985q,sklyanin1988boundary,pasquier1990common,gomez1996quantum}.
  {While the isotropic XXX model corresponds to the $q\to 1$ limit, the XXZ chain with {specific} open boundary conditions is specifically characterized by its $U_{q}(\mathfrak{su}(2))$ invariance, which ensures the existence of a set of conserved quantities that commute with the Hamiltonian}~\cite{kulish1991general}.
  Various other long-range interacting models with $U_{q}(\mathfrak{su}(2))$
  symmetry exist, including the braid-translated $q$-deformed periodic XXZ chain~\cite{martin1993algebraic,martin1994blob},
  the $q$-deformed Haldane--Shastry model~\cite{lamers2022spin}, and systems with
  quantum affine algebra symmetry~\cite{hakobyan1996spin}. Elliptic extensions
  and integrable deformations have also been constructed~\cite{matushko2022elliptic,klabbers2024deformed}.
  These $q$-deformed models typically exhibit nonlocal interactions, which is a
  feature shared by the model introduced here.

  More generally, the Hopf algebra formalism~\cite{majid2000foundations} and the
  $\mathfrak{su}_{q}(2)$ deformation~\cite{biedenharn1995quantum,curtright1991quantum}
  have led to numerous physical applications, including deformed bosons in
  pairing correlations~\cite{bonatsos1999quantum}, perturbative studies of the Dicke
  model~\cite{ballesteros1999spectrum}, the $q$-rotator in diatomic molecular
  spectroscopy~\cite{chang1991suq}, thermodynamic properties of deformed fermionic
  systems~\cite{ubriaco1996thermodynamics,ubriaco1998lambda}, and stability
  analyses comparing $\mathfrak{su}(2)$ and $\mathfrak{su}_{q}(2)$ symmetric
  systems~\cite{ubriaco2012geometry}. This approach also permits analytical solutions,
  as in models where bosonic interactions can be embedded via an appropriate $q$-deformation
  of the $\mathfrak{su}(2)$ fermionic algebra~\cite{ballesteros2002fermion}.

  In this work, we consider the $q$-deformation of the KS model introduced in \cite{Ballesteros:2025cia}.
  This is an integrable model that depends on the deformation parameter $q$ with
  the underlying $\mathfrak{su}_{q}(2)$ symmetry. The Curie temperature was obtained
  in \cite{Ballesteros:2025cia} for the deformed case, but the thermodynamic
  properties of this deformed model were not discussed. This is the aim of this paper.
  In particular, we will focus on the specific heat, magnetic susceptibility, and
  magnetisation when the angular momentum of the individual spins is $j=1/2$,
  for both the antiferromagnetic and ferromagnetic scenarios. The deformation parameter
  modifies the energy spectrum by increasing the spacing between the energy levels,
  thereby shifting the thermodynamic behaviour toward higher temperatures. In the
  ferromagnetic case, the deformation shifts the Curie temperature to higher
  values and narrows the specific heat peaks, whereas in the antiferromagnetic case,
  analytical approximations based on the two most probable energy levels
  accurately describe the thermodynamic quantities and the phase transitions. Overall,
  deformation introduces non-identical spin couplings, which significantly influence
  the thermodynamic behaviour compared to that of the undeformed KS model. 

{Moreover, we discuss the effect in thermodynamic magnitudes of complex $q$, being on the unit circle but not a root of unity. This restriction is important because, at
roots of unity, the representation theory of $U_q(\mathfrak{su}(2))$ changes
substantially: finite-dimensional representations may become reducible or
indecomposable, truncation effects can appear, and the thermodynamic
interpretation of the deformation is no longer continuously connected to the
generic case \cite{biedenharn1995quantum}. For $|q|=1$, the deformation introduces an oscillatory redistribution of spectral weight among spin multiplets rather than the exponential amplification characteristic of real $q$. Since the analytical derivations obtained for the real case are automatically valid for complex $q$ (not a root of unity), we mainly focus on the graphical representation of eigenvalues and thermodynamic magnitudes for $N=4$, although some qualitative explanations are given. We further exploit this complex $q$-deformation to establish a connection with the physics of geometrically frustrated quantum magnets. Specifically, we compare the thermodynamic predictions of the $q$-deformed Kittel--Shore model with exact-diagonalisation results for the isolated tetrahedral Hamiltonian describing the intratetrahedral interactions in the Cu$_2$Te$_2$O$_5$X$_2$ ($X=$ Cl, Br) family of compounds~\cite{jaglicic2006magnetic}. The comparison reveals that the deformation phase $\theta$ (with $q=e^{i\theta}$) acts as an effective algebraic parameter that encodes the microscopic frustration arising from the non-equivalent exchange couplings within the distorted tetrahedra, establishing a thermodynamic equivalence $(J_1,J_2)\to (I,\theta)$ between the microscopic and the $q$-deformed descriptions.}

  The structure of this paper is organized as follows. In Sec. \ref{sec:ksk},
  the Kittel-Shore model and its deformation based on $\mathfrak{su}_{q}(2)$
  symmetry are presented. Then, in Sec.~\ref{sec:kskferro}, we analyse the
  thermodynamic properties in the undeformed and deformed scenarios for the
  ferromagnetic case, whereas Sec.~\ref{sec:kskanti} focuses on the
  antiferromagnetic case. {{Sec.}~\ref{sec:complex} analyses the complex $q$-deformation ($|q|=1$, not a root of unity) and its application to frustrated tetrahedral quantum magnets, establishing a thermodynamic equivalence between the microscopic Hamiltonian of Cu$_2$Te$_2$O$_5$X$_2$ compounds and the $q$-deformed effective model.} Finally, Sec.~\ref{sec:conclusions} presents our conclusions,
  highlighting the potential applications of this deformed model and directions
  for future research.

  \section{The Kittel--Shore model and its \texorpdfstring{$q$}{q}-deformation}
  \label{sec:ksk} We present the KS model~\cite{kittel1965development} for a
  system of $N$ spins with an external magnetic field, but we start from the expression
  given in~\cite{al1998exact} (a difference of a factor of 2), making it easier
  to compare that paper and our results. This Hamiltonian reads
  \begin{equation}
    H_{KS}=- I \sum_{i<j}^{N}\Vec{J}^{{(i)}}\cdot\Vec{J}^{{(j)}}-\gamma h\sum_{{k}=1}
    ^{N}J_{z}^{({k})}, \label{eq:hamiltonian1}
  \end{equation}
  where $I$ is the interaction constant between the spins, $\gamma=g\mu_{B}$ in
  the usual notation, and $h$ is the external magnetic field. The interaction
  constant distinguishes between antiferromagnetic and ferromagnetic cases
  depending on its sign (negative or positive, respectively).

  By definition, the angular momenta operators are given by
  \begin{equation}
    J_{x}= \frac{1}{2}\sigma_{x}, \qquad J_{y}= \frac{1}{2}\sigma_{y}, \qquad J_{z}
    = \frac{1}{2}\sigma_{z}, \label{eq:angular}
  \end{equation}
  being $\sigma_{i}$ the Pauli matrices
  \begin{equation}
    \sigma_{x}=
    \begin{pmatrix}
      0 & 1 \\
      1 & 0 \\
    \end{pmatrix}, \qquad \sigma_{y}=
    \begin{pmatrix}
      0 & -i \\
      i & 0  \\
    \end{pmatrix}, \qquad \sigma_{z}=
    \begin{pmatrix}
      1 & 0  \\
      0 & -1 \\
    \end{pmatrix}. \label{eq:pauli}
  \end{equation}
  With the definition of angular momentum operators given in~\eqref{eq:angular},
  we can write
  \begin{equation}
    J_{\pm}= J_{x}\pm i J_{y},
  \end{equation}
  and, therefore
  \begin{equation}
    J_{x}= \frac{1}{2}(J_{+}+ J_{-}), \qquad J_{y}= - \frac{i}{2}(J_{+}- J_{-}),
  \end{equation}
  so the matrix form of the $\{J_{+},J_{-},J_{z}\}$ operators are given by
  \begin{equation}
    J_{+}=
    \begin{pmatrix}
      0 & 1 \\
      0 & 0 \\
    \end{pmatrix}, \qquad J_{-}=
    \begin{pmatrix}
      0 & 0 \\
      1 & 0 \\
    \end{pmatrix}, \qquad J_{z}= \frac{1}{2}
    \begin{pmatrix}
      1 & 0  \\
      0 & -1 \\
    \end{pmatrix}. \label{eq:jmp}
  \end{equation}

  Then, the Hamiltonian of Eq.~\eqref{eq:hamiltonian1} can be written as \cite{Ballesteros:2025cia}
  \begin{align}
    H_{KS} & = -\frac{I}{2}\left( \sum_{i=1}^{N}J_{-}^{(i)}J_{+}^{(i)}+ \sum_{{w}=1}^{N-1}\sum_{r={w}+1}^{N}\left( J_{-}^{({w})}J_{+}^{(r)}+ J_{+}^{({w})}J_{-}^{(r)}\right) + \sum_{{l}=1}^{N}J_{z}^{({l})}\sum_{s=1}^{N}\left(J_{z}^{(s)}+ {\mathbb{I}}\right)- \sum_{{t}=1}^{N}C^{({t})}\right) -\gamma h \sum_{{k}=1}^{N}J_{z}^{({k})}, \label{finalKSpm}
  \end{align}
  where $C^{(i)}$ is the Casimir of the $\mathfrak{su}(2)$ algebra, given by
  \begin{equation}
    C^{(i)}= J_{-}^{(i)}J_{+}^{(i)}+J_{z}^{(i)}(J_{z}^{(i)}+{\mathbb{I}})=J_{+}^{(i)}J_{-}^{(i)}
    + J_{z}^{(i)}(J_{z}^{(i)}-{\mathbb{I}}).
  \end{equation}
  Consequently, the eigenvalues of the Hamiltonian~\eqref{finalKSpm} for a spin
  chain consisting of particles with identical spin $j$ are
  \begin{equation}
    E_{N{,}J{,}m}=-\frac{I}{2}\left(J(J+1)-N j(j+1)\right)-\gamma h m, \label{eq:energy}
  \end{equation}
  where $J=\sum_{i=1}^{N}j=N j$ and $m=\sum_{i=1}^{N}m_{i}$, being $m_{i}$ is the
  eigenvalue of the operator $J_{i}^{z}$. This implies that the eigenvectors of
  the Hamiltonian~\eqref{finalKSpm} are those given by the Clebsch--Gordan
  coefficients.

  The Hamiltonian~\eqref{finalKSpm} can be deformed owing to its coalgebra symmetry~\cite{ballesteros1998systematic}. {With the exception of the $j=1/2$ case, the generators become $q$-dependent, so for the generic deformed case we switch to the algebra $\mathfrak{su}_{q}(2)$ basis $\{\bar{J}_+, \bar{J}_-, \bar{J}_z\}$. These generators satisfy the commutation relations (see \cite{biedenharn1995quantum}):}
  \begin{equation}
    [\bar{J}_z,\bar{J}_\pm]= \pm \bar{J}_\pm,\qquad [\bar{J}_+,\bar{J}_-]= \left[ 2 \bar{J}_z\right]_q = \frac{q^{\bar{J}_z}-q^{-\bar{J}_z}}{q^{1/2}-q^{-1/2}} ,
    \label{suq2}
\end{equation}
    {where  we have made use of the $q$-symbol}
  \begin{equation}
    [n]_{q}\coloneqq \frac{q^{n/2}-q^{-n/2}}{q^{1/2}-q^{-1/2}}. \label{eq:qnumber}
  \end{equation}
  Then, a new superintegrable model (maximally superintegrable for vanishing
  external magnetic field) is obtained when deforming this Hamiltonian through $U
  _{q}(\mathfrak{su}(2))$, which reads \cite{Ballesteros:2025cia}
  \begin{align}
    \tilde H_{KS}^{q}= & -\frac{I}{2}\left(\sum_{i=1}^{N}\text{ exp}\left[-\eta \sum_{j=1}^{i-1}\bar{J}_{z}^{(j)}\right]\bar{J}_{-}^{(i)}\bar{J}_{+}^{(i)}\text{ exp}\left[\eta\sum_{h=i+1}^{N}\bar{J}_{z}^{(h)}\right]+\sum_{{w}=1}^{N-1}\sum_{r={w}+1}^{N}\left(e^{\eta/2}\bar{J}_{-}^{({w})}\bar{J}_{+}^{(r)}+e^{-\eta/2}\bar{J}_{+}^{({w})}\bar{J}_{-}^{(r)}\right)\cdot\right. \notag                 \\
    & \cdot\text{ exp}\left[-\eta \frac{\bar{J}_{z}^{({w})}}{2}\right]\cdot\text{ exp}\left[\eta \frac{\bar{J}_{z}^{(r)}}{2}\right]\cdot\prod_{t=1}^{{w}-1}\text{ exp}\left[-\eta \bar{J}_{z}^{(t)}\right]\prod_{k=r+1}^{N}\text{ exp}\left[\eta \bar{J}_{z}^{(k)}\right]+\left[\sum_{{l}=1}^{N}\bar{J}_{z}^{({l})}\right]_{q}\left[\sum_{s=1}^{N}\bar{J}_{z}^{(s)}+{\mathbb{I}}\right]_{q}- \notag \\
    & -\sum_{{a}=1}^{N}C_{q}^{({a})}\Biggl)-\gamma h \sum_{{o}=1}^{N}\bar{J}_{z}^{({o})}, \label{finalqKSpm}
  \end{align}
  where $q=e^{\eta}$ {and} $C_{q}^{(i)}$ is the Casimir operator for the algebra $\mathfrak{su}
  _{q}(2)$
  \begin{equation}
    C_{q}^{(i)}= \bar{J}_{-}^{(i)}\bar{J}_{+}^{(i)}+ [\bar{J}_{z}^{(i)}]_{q}[\bar{J}_{z}^{(i)}+{\mathbb{I}}
    ]_{q}=\bar{J}_{+}^{(i)}\bar{J}_{-}^{(i)}+ [\bar{J}_{z}^{(i)}]_{q}[\bar{J}_{z}^{(i)}-{\mathbb{I}}]_{q}.
    \label{qcas}
  \end{equation}

  Then, for $q=1$ (equivalently, $\eta=0$), the deformed Hamiltonian~\eqref{finalqKSpm}
  leads to the undeformed KS Hamiltonian~\eqref{finalKSpm}. For the particular case
  of {a single $1/2$-spin particle}, the angular momentum operators of the algebra
  $\mathfrak{su}_{q}(2)$ are those in Eq.~\eqref{eq:jmp}, that is, independent of
  the parameter $q$ \cite{biedenharn1995quantum}.

  Following the rules of action of angular momentum operators, it is easy to
  obtain that the corresponding eigenvalues of this deformed Hamiltonian are
  \begin{equation}
    E^{q}_{N{,}J{,}m}=-\frac{I}{2}\left([J]_{q}[J+1]_{q}-N [j]_{q}[j+1]_{q}\right)-\gamma
    h m. \label{eq:qenergy}
  \end{equation}
  The energy distribution of the deformed model reveals that deformation
  increases the spacing between the energy levels \cite{Ballesteros:2025cia}. This
  dilation of the differences between levels is more prominent for those with a larger
  $J$.

  The distribution of energy levels in the undeformed ferromagnetic and antiferromagnetic
  cases was studied in~\cite{czachor2008energy,Ballesteros:2025cia}. In the
  ferromagnetic case, the ground energy states are the most separated energy levels
  and have the lowest degree of degeneracy. In contrast, the lowest energy levels
  are very close in the antiferromagnetic case. This allows the energy
  distribution to be considered as a continuum in this range. Moreover, the
  antiferromagnetic degeneracy is larger in the low-energy range. These phenomena
  are mainly responsible for the fact that we can use good (approximate) analytical
  approaches for the antiferromagnetic case but not for the ferromagnetic case,
  as discussed in the following sections.

  Once the Hamiltonian is given, the partition function for the $N$-spins problem
  can be obtained from
  \begin{equation}
    Z_{N}=\sum_{J}\sum_{m=-J}^{m=J}d_{N{,}J}\, \text{exp}(-\beta E_{N{,}J{,}m}),
  \end{equation}
  where $\beta=1/k_{B}T$ and the coefficients $d_{NJ}$ represent the multiplicity
  of the state of $N$ spins with total angular momentum $J$ and quantum number $m$.
  These coefficients are given by~\cite{morse1932theory}
  \begin{equation}
    d_{N{,}J}=\Omega(N,J)-\Omega(N,J+1), \quad \text{with}\quad \Omega(N,J)=\text{coefficient
    of }x^{J}\text{ in }\left(x^{j}+x^{j-1}+\cdots +x^{-j}\right)^{N}.
  \end{equation}
  For the particular case considered in this work, $j=1/2$, the previous
  expression can be written as \cite{al1998exact}
  \begin{equation}
    d_{N{,}J}= \binom{N}{N/2-J}-\binom{N}{N/2-(J+1)}. \label{eq:deg}
  \end{equation}
  For the deformed case, we will use
  \begin{equation}
    Z_{N}^{q}=\sum_{J}\sum_{m=-J}^{m=+J}d_{N{,}J}\, \text{exp}(-\beta E^{q}_{N{,}J{,}m}),
    \label{partition}
  \end{equation}
  {Notably, for generic values (not roots of unity) of $q$ the degeneracies of this model remain the same as in the undeformed case.}

  This partition function can be used to obtain the Helmholtz free energy,
  $F=-(k_{B}T/N)\log(Z)$. All thermodynamic functions can be obtained by differentiating
  this energy. The specific heat, magnetic susceptibility, and magnetisation are
  given by
  \begin{align}
    C_{V} & =-T\left.\frac{\partial^{2}F}{\partial T^{2}}\right|_{h=0},\qquad \chi=-\left.\frac{\partial^{2}F}{\partial H^{2}}\right|_{h=0},\qquad M=-\frac{\partial F}{\partial H}. \label{eq:thermody}
  \end{align}
  In the following, we study these properties for antiferromagnets and ferromagnets
  corresponding to $I=-1$ and $I=1$, respectively. {Before going on, we would like to remark that the cases $N=2$ and $N=3$ of the deformed scenario do not differ from the undeformed one, since the terms proportional to the deformation parameter can be absorbed in the coupling constant. }

\subsection{{Invariance of the model under the transformation \texorpdfstring{$q\rightarrow q^{-1}$}{q->q-1}}}
\label{invariance}

{The fundamental building block of $U_q(\mathfrak{su}(2))$, the $q$-deformed integer (or $q$-number), transforms under parameter inversion as:
\begin{equation}
    [n]_{q^{-1}} = \frac{(q^{-1})^{n/2} - (q^{-1})^{-n/2}}{(q^{-1})^{1/2} - (q^{-1})^{-1/2}} = \frac{q^{-n/2} - q^{n/2}}{q^{-1/2} - q^{1/2}}= \frac{-(q^{n/2} - q^{-n/2})}{-(q^{1/2} - q^{-1/2})} = \frac{q^{n/2} - q^{-n/2}}{q^{1/2} - q^{-1/2}} = [n]_q.
\end{equation}}
{Consequently, the identity $[n]_{q^{-1}} = [n]_q$ holds universally for all $n \in \mathbb{C}$. Applying this symmetry directly to the energy spectrum yields:
\begin{equation}
    E_{N,J,m}^{q^{-1}} = -\frac{I}{2}\left( [J]_{q^{-1}} [J+1]_{q^{-1}} - N [j]_{q^{-1}} [j+1]_{q^{-1}} \right) - \gamma h m = E_{N,J,m}^{q}.
\end{equation}}
{Since the partition function $Z_{N}^{q}$ and the degeneracy $d_{N,J}$ depend solely on the eigenvalues $E_{N,J,m}^{q}$, Eq. (\ref{partition}), it follows that $Z_N^{q^{-1}} = Z_N^q$. Consequently, all derived thermodynamic quantities (Helmholtz free energy $F$, specific heat $C_{V}$, magnetic susceptibility $\chi$, and magnetization $M$) are exactly identical under the inversion $q \to q^{-1}$.}

{At the algebraic level, an analogous invariance holds for the Casimir operator of each site $i$:
\begin{equation}
    C_{q^{-1}}^{(i)} = J_-^{(i)} J_+^{(i)} + [J_z^{(i)}]_{q^{-1}} [J_z^{(i)} + \mathbb{I}]_{q^{-1}}= J_-^{(i)} J_+^{(i)} + [J_z^{(i)}]_{q} [J_z^{(i)} + \mathbb{I}]_{q}=C_{q}^{(i)}.
\end{equation}}
{Furthermore, setting $q = e^\eta$, the parameter inversion $q \to q^{-1}$ corresponds to $\eta \to -\eta$. In Eq.~(\ref{finalqKSpm}), this sign change flips the sign of the exponents in the operator factors $\exp[\eta J_z^{(k)}]$. Algebraically, this transformation is equivalent to a site-reversal automorphism $i \to N - i + 1$ of the Hopf algebra $U_q(\mathfrak{su}(2))$, or equivalently, to a unitary conjugation via a parity-like operator. Because unitary transformations preserve both the energy spectrum and degeneracies in finite-dimensional representations (provided $q$ is not a non-trivial root of unity), the Hamiltonian parameterized by $q^{-1}$ is unitarily equivalent to that parameterized by $q$.}

{In summary, the model exhibits exact invariance under the transformation $q \to q^{-1}$ across both its underlying algebraic structure and its thermodynamic behavior.}

  \section{The ferromagnetic case}
  \label{sec:kskferro} We begin with the {undeformed ($q=1$)} ferromagnetic case
  and discuss the KS model for a small number of spins and for the thermodynamic
  limit. {Subsequently, the same analysis will be carried out for the deformed case.}

  \subsection{Small number of spins}
\begin{figure}[H]
    \centering
    \includegraphics[scale=0.9]{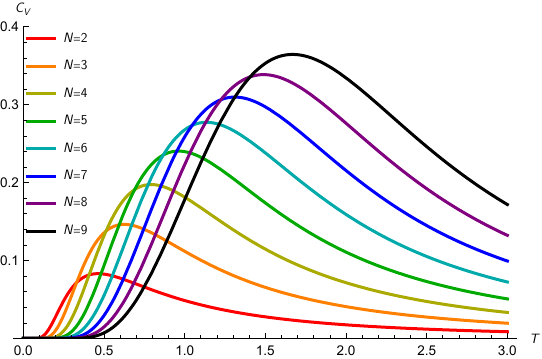}
    \caption{Specific heat as a function of temperature for values of the number
    of particles $N=2$ (red), $3$ (orange), $4$ (yellow), $5$ (green), $6$ (cyan),
    $7$ (blue), $8$ (purple), and $9$ (black) for the ferromagnetic case.}
    \label{kskj1/2cvf}
  \end{figure}

  Using the expressions in Eq.~\eqref{eq:thermody}, the thermodynamic properties
  of the undeformed model are {displayed using numerical computation}. As
  illustrated in Fig.~\ref{kskj1/2cvf}, the specific heat exhibits a higher
  maximum value and is shifted to the right for larger values of $N$. Moreover,
  we observe the same behaviour regardless of whether $N$ is odd or even.

We can now try to understand why the specific heat shows this behaviour, in particular, why it increases and its maximum is shifted to higher temperatures as $N$ increases. The density of the energy levels is considerably separated in the ferromagnetic case; therefore, its energy profile is discrete throughout the range considered~\cite{czachor2008energy}. Moreover, it can be seen that an important contribution to the peak is due to the two most probable levels (the lower the number of particles, the greater the contribution). Therefore, these temperatures are approximately proportional to the difference between the {ground-state} and first excited energy levels, $T\propto\Delta E$. A simple calculation using Eq.~\eqref{eq:energy} yields $E_{1}-E_{0}=I N/2$, which means $T\propto N$. {This behavior reveals that the unscaled ferromagnetic model lacks extensivity, as the energy scales required to observe thermal effects grow with the system size. For large $N$, this leads to physically unrealistic temperature regimes, which is why the analysis for $N > 9$ is not presented in this unscaled context. However, as discussed later in this section, the physical consistency of the model is recovered in the thermodynamic limit by introducing the Kac scaling \cite{kac1963van}, $I \to I/N$. This rescaling ensures a well-defined limit where the energy per particle remains finite, allowing for a complete thermodynamic analysis of the system.}

\begin{figure}[H]
    \centering
    \includegraphics[scale=0.9]{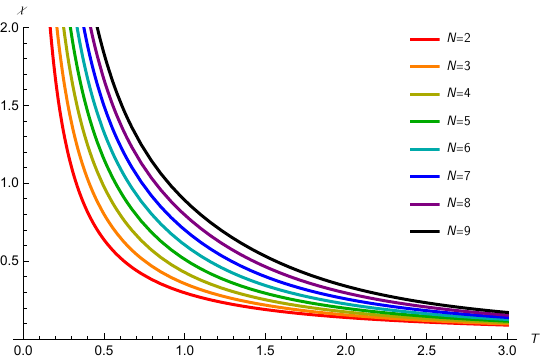}
    \caption{Magnetic susceptibility as a function of temperature for values of
    the number of particles $N=2$ (red), $3$ (orange), $4$ (yellow), $5$ (green),
    $6$ (cyan), $7$ (blue), $8$ (purple), and $9$ (black) for the ferromagnetic case.}
    \label{kskj1/2xf}
  \end{figure}

For magnetic susceptibility, one can observe a behaviour $1/(NT)$, independent of the parity of $N$, as shown in Fig.~\ref{kskj1/2xf}.

\begin{figure}[H]
    \centering
    \includegraphics[scale=0.9]{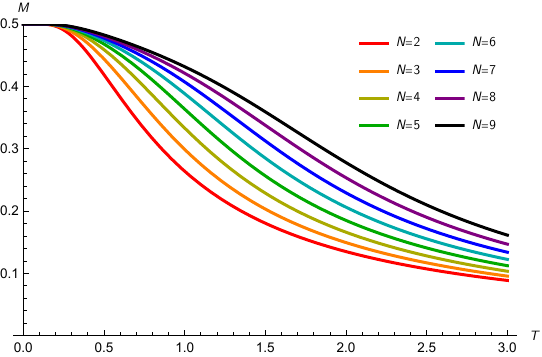}
    \caption{Magnetisation as a function of temperature for values of the number of particles $N=2$ (red), $3$ (orange), $4$ (yellow), $5$ (green), $6$ (cyan), $7$ (blue), $8$ (purple), and $9$ (black), with ${h\gamma=1}$ for the ferromagnetic  case.}
    \label{kskj1/2mf}
  \end{figure}

Finally, the magnetisation is shown in Fig.~\ref{kskj1/2mf} for an external magnetic field (in ${h\gamma=1}$ units). We observe that at zero temperature, the magnetisation is $1/2$, which is independent of the number of spins. Because all spins point in the same direction in this case, the magnetisation is given by $M=m/N=N j/N =1/2$. {At higher temperatures, the spins become randomly oriented due to thermal agitation, leading to a vanishing net magnetisation.}

\begin{figure}[H]
    \centering
    \includegraphics[scale=0.9]{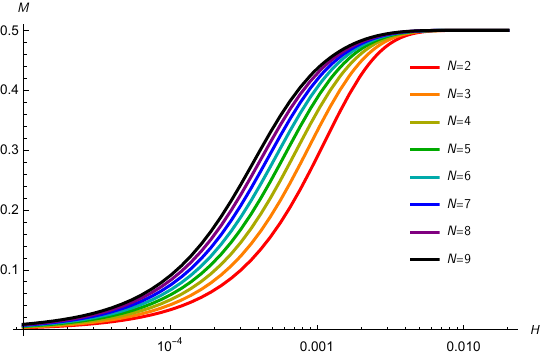}
    \caption{Magnetisation as a function of magnetic field for values of the number of particles $N=2$ (red), $3$ (orange), $4$ (yellow), $5$ (green), $6$ (cyan), $7$ (blue), $8$ (purple), and $9$ (black) for temperature $T={0.001}$, {with $h{\gamma}=0$} for the ferromagnetic case.}
    \label{ferrotransition}
  \end{figure}

We also find a critical change of the magnetisation at $h{\gamma}=0$ for the temperature $T=0^{+}$ obtained in \cite{kac1968statistical}, as shown in Fig.~\ref{ferrotransition}.
The chosen temperature is $T=0.001$ and the closer this value is to 0, the {more similar the curves become to a step function in the $T\to0$ limit}. The least energetic state for $h=0$ is shared by all configurations with the largest value $J$. Therefore, the average magnetisation of the set of configurations is zero. When a small magnetic field is applied, this degeneracy is broken, and the least energetic state is given by all spins up with the largest $J$. Thus, the magnetisation reaches its maximum value at this point and does not decrease again because as the magnetic field increases, the previous state becomes less energetic. This occurs for $N$ being even and odd.

\subsection{Thermodynamic limit}
To consider the thermodynamic limit, we change $I\rightarrow I/N$, allowing the thermodynamic properties to be accurately described \cite{czachor2008energy}.

\subsubsection{Thermodynamic properties}
Now, the behaviour of the thermodynamic properties are examined in the thermodynamic limit {by numerical evaluation of the exact partition function}. The numbers of particles $N=100,500$ and $1000$ are selected because parity does not play a significant role in the ferromagnetic case. This is due to the fact that, unlike in the antiferromagnetic case, where the ground state depends on the parity of $N$ (as we will see in the following section), the ferromagnetic ground states ($J=N/2,N/2-1$) remain unchanged.

\begin{figure}[H]
    \centering
    \includegraphics[scale=0.9]{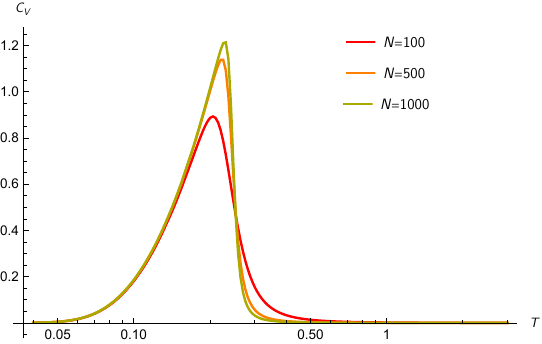}
    \caption{Specific heat as a function of temperature for values of the number
    of particles $N=100$ (red), $500$ (orange), and $1000$ (yellow), for the interaction
    constant $I=1/N$ ferromagnetic case.}
    \label{cvlimitferro}
  \end{figure}

The three properties analysed here diverge or show peaks near the Curie temperature ($T_{C}=0.25$ \cite{kittel1965development}), indicating a phase transition \cite{kittel2018introduction}. In Fig. \ref{cvlimitferro}, we observe that the specific heat peaks occur at nearly the same temperature and become slightly larger as the number of particles increases, showing the transition.

\begin{figure}[H]
    \centering
    \includegraphics[scale=0.9]{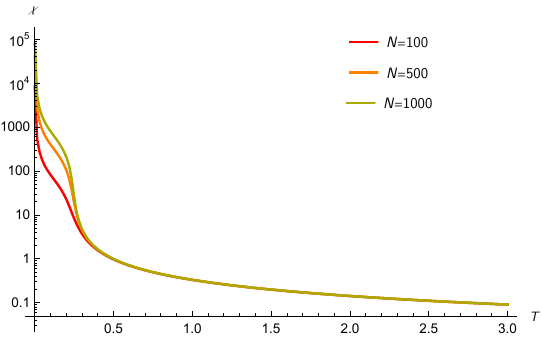}
    \caption{Magnetic susceptibility as a function of temperature for values of
    the number of particles $N=100$ (red), $500$ (orange), and $1000$ (yellow), for
    the interaction constant $I=1/N$ ferromagnetic case.}
    \label{chilimitferro}
  \end{figure}

In Fig. \ref{chilimitferro}, magnetic susceptibility exhibits convergence at both low temperatures, where it diverges to infinity, and at high temperatures, where it approaches zero. In the intermediate range, the curve shows a Curie transition, which will be discussed later. The curves differ only during this transition, which occurs at almost the same temperature but reaches higher susceptibility values as the number of particles increases.

\begin{figure}[H]
    \begin{minipage}{0.4\textwidth}
      (c)
      \includegraphics[scale=0.7]{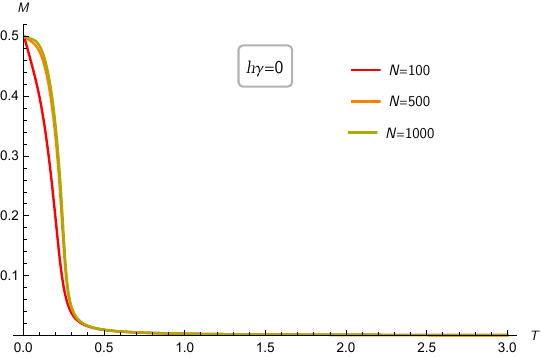}
    \end{minipage}
    \hspace{15mm}
    \begin{minipage}{0.4\textwidth}
      (d)
      \includegraphics[scale=0.7]{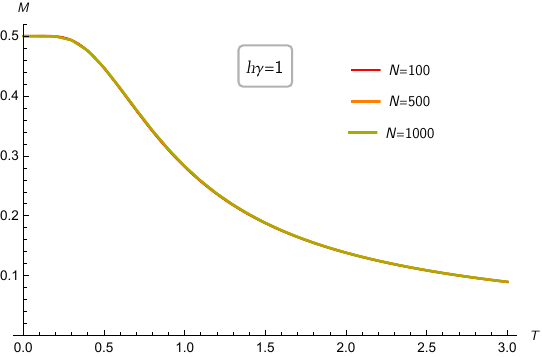}
    \end{minipage}
    \caption{Magnetisation as a function of temperature for values of the number of particles $N=100$ (red), $500$ (orange), and $1000$ (yellow), for the interaction constant $I=1/N$ and (a) external magnetic field {$h\gamma=0$}, (b) {$h\gamma=1$} ferromagnetic case. They all are coincident.}
    \label{mlimitferro}
  \end{figure}

In Fig. \ref{mlimitferro}, we present the magnetisation with and without an external magnetic field. This thermodynamic property is directly related to magnetic susceptibility; therefore, the one without an external magnetic field also allows one to observe the Curie transition. In the figure with an external magnetic field, we can see that all the $N$ curves coincide in the thermodynamic limit, independently of the temperature.
In the absence of an external magnetic field, we can observe from the thermodynamic properties that the Curie temperature is approximately the same for every number of particles, as discussed in~\cite{kittel1965development}. When an external magnetic field is introduced, differences appear. The modification of the coupling and the presence of an external magnetic field cause the magnetic term in Eq. (\ref{eq:energy}) to become dominant. Consequently, the most probable state remains $\ket{N/2, N/2}$, but the energy distribution undergoes significant changes because the energy differences are proportional to the external magnetic field $h$. Thus, the magnetisation takes the value $M=0.5$ at $T=0$ and decreases when the next energy level is reached. This decrease is practically the same for all curves because the energy levels depend on $N$, and the numbers chosen here are sufficiently large that the differences in magnetisation are minimal.

{By the dominance of the magnetic part, it is possible to obtain an analytical approximation for the magnetisation for large $N$ by considering the most probable levels ($J=N/2, N/2-1$):}
 \begin{equation}
    M^{\text{approx}}_{\text{ferro}}{\approx}\frac{A-B+C+D}{E}, \label{eq:magferro}
  \end{equation}
  where
  \begin{equation*}
    A=(N-2) (N-1) e^{1/T},
  \end{equation*}
  \begin{equation*}
    B=N (N-1) e^{2/T},
  \end{equation*}
  \begin{equation*}
    C=N e^{\frac{1}{2 T}}-(N+2) e^{\frac{3}{2 T}},
  \end{equation*}
  \begin{equation*}
    D=e^{N/T}\left((N-2) (N-1) \left(-e^{1/T}\right)-N e^{\frac{5}{2 T}}+(N+2) e^{\frac{3}{2
    T}}+N (N-1)\right),
  \end{equation*}
  \begin{equation*}
    E=2 N \left(e^{1/T}-1\right) \left((N-1) e^{1/T}-e^{N/T}\left(N+e^{\frac{3}{2
    T}}-1\right)+e^{\frac{1}{2 T}}\right).
  \end{equation*}

This expression does not accurately reproduce the magnetisation at temperatures above the Curie transition because the ground states are the least degenerate. However, it provides a highly precise approximation at low temperatures, particularly for identifying transition points. In the undeformed case, an exact analytical expression for the thermodynamic functions of the model, including the magnetisation, can be obtained without approximations, as shown in Ref.~\cite{czachor2002verification}. However, such a closed-form treatment is not feasible for the $q$-deformed Hamiltonian, so here we employ the most-probable-level approximation $(J = N/2,N/2-1)$ in the thermodynamic limit to obtain an {approximate} analytic expression {(Eq. (\ref{eq:magferro}))} that will serve as a reference for comparison with the deformed case.

\subsection{\texorpdfstring{$q$}{q}-deformation for small number of spins}
We begin by considering the $q$-deformed model for a small number of spins. {As in the undeformed case, all the thermodynamic quantities presented in this subsection are obtained by direct numerical evaluation of the partition function~\eqref{partition}.} We can note that the modification due to the variation of the deformation parameter becomes more appreciable for $q$ and $N$ larger. This is due to the fact that the energy values depend on the number of particles and $q$-number properties, so that the modulus of the energy of many particles becomes greater than the one of a few particles when a deformation is introduced ($[|x|+1]_{q}-(|x|+1)>[ |x|]_{q}-|x|$).

\begin{figure}[H]
    \begin{minipage}{0.4\textwidth}
      (a)
      \includegraphics[scale=0.7]{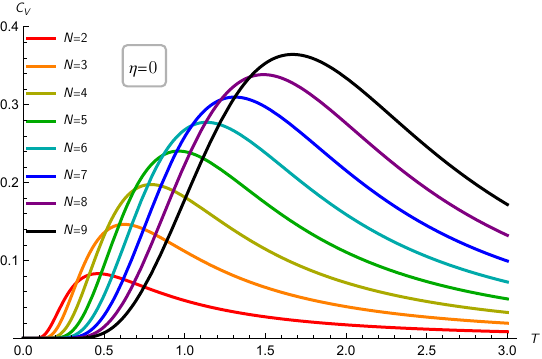}
    \end{minipage}
    \hspace{15mm}
    \begin{minipage}{0.4\textwidth}
      (b)
      \includegraphics[scale=0.7]{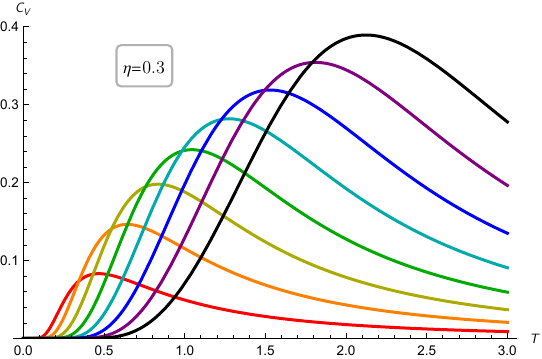}
    \end{minipage}
    \caption{Specific heat as a function of temperature for values of the number
    of particles $N=2$ (red), $3$ (orange), $4$ (yellow), $5$ (green), $6$ (cyan),
    $7$ (blue), $8$ (purple), and $9$ (black) and of the parameter (a) $\eta=0$
    and (b) $\eta=0.3$, for the ferromagnetic case.}
    \label{qkskj1/2cvfd}
  \end{figure}

We begin with the specific heat shown in Fig.~\ref{qkskj1/2cvfd}. It is important to note that the maxima remain with the same magnitude and are placed further to the right for increasing values of $\eta$, which follows the reasoning discussed above. Note that the negative values of the parameter $\eta$ have not been plotted because the model is symmetric with respect to $\eta=0$.

  As for the undeformed case, the temperature at which the maximum specific heat
  appears is proportional to the difference between the two {ground} levels,
  $T \propto E_{1}-E_{2}$. 
  Therefore, the maximum value of the deformation parameter considered was $\eta=
  0.3$. Higher deformation parameters do not allow us to observe the peaks
  properly within the considered temperature range.

\begin{figure}[H]
    \begin{minipage}{0.4\textwidth}
      (a)
      \includegraphics[scale=0.7]{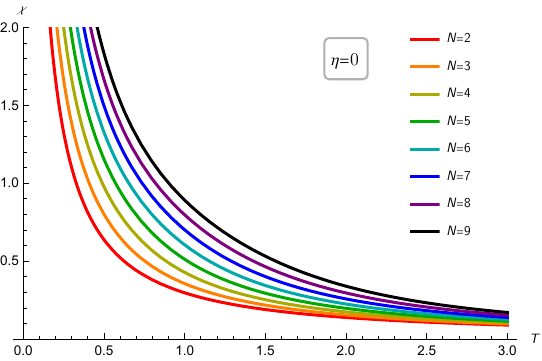}
    \end{minipage}
    \hspace{15mm}
    \begin{minipage}{0.4\textwidth}
      (b)
      \includegraphics[scale=0.7]{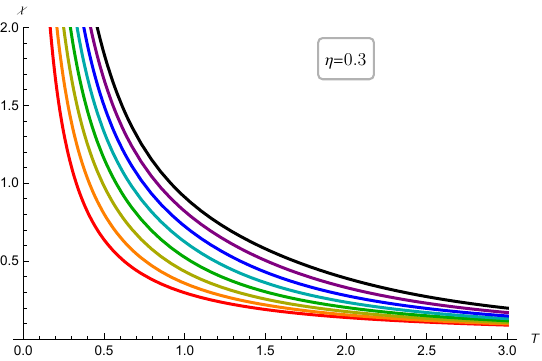}
    \end{minipage}
    \caption{Magnetic susceptibility as a function of temperature for values of
    the number of particles $N=2$ (red), $3$ (orange), $4$ (yellow), $5$ (green),
    $6$ (cyan), $7$ (blue), $8$ (purple), and $9$ (black) and of the parameter (a)
    $\eta=0$ and (b) $\eta=0.3$, for the ferromagnetic
    case.}
    \label{qkskj1/2xf}
  \end{figure}

  The magnetic susceptibility is shown in Fig.~\ref{qkskj1/2xf}. We observe that
  while there are small differences between the plots depending on the value of
  $\eta$, the behaviour $1/(NT)$ is not modified at high temperatures.

\begin{figure}[H]
    \begin{minipage}{0.4\textwidth}
      (a)
      \includegraphics[scale=0.7]{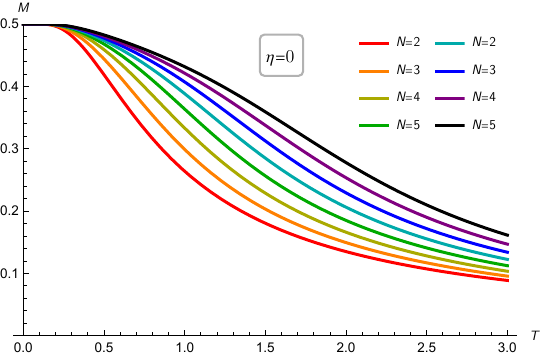}
    \end{minipage}
    \hspace{15mm}
    \begin{minipage}{0.4\textwidth}
      (b)
      \includegraphics[scale=0.7]{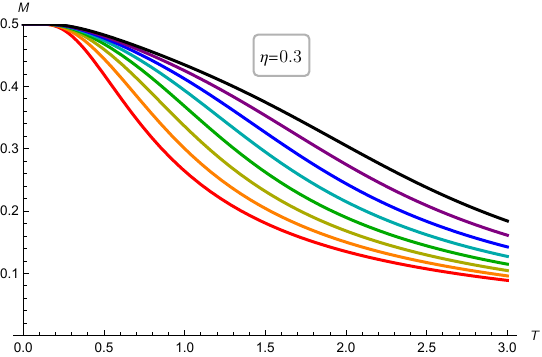}
    \end{minipage}
    \caption{Magnetisation as a function of temperature for values of the number
    of particles $N=2$ (red), $3$ (orange), $4$ (yellow), $5$ (green), $6$ (cyan),
    $7$ (blue), $8$ (purple), and $9$ (black) and of the parameter (a) $\eta=0$
    and (b) $\eta=0.3$, for the ${h\gamma=1}$ ferromagnetic
    case.}
    \label{qkskj1/2mf}
  \end{figure}

  The magnetisation is shown in Fig.~\ref{qkskj1/2mf}. Here, we do not find any difference
  between $N$ odd or even, as in the undeformed case. The only modification with
  respect to the undeformed case is a small change in the convexity of the function
  for large $N$ when $\eta$ increases, because the $q$-deformation makes the
  energy levels become more separated, and therefore the interaction energy is greater
  with respect to the undeformed case.

As in the undeformed case, we find a change in the magnetisation for $h=0$ at $T=0^{+}$, independent of the value of $q$; therefore, the corresponding graphs for different values of $q$ are not presented. Furthermore, at $h=0$, the state with the lowest energy is shared by every possible configuration (each $m$) at the highest $J$ level for any $q$, given a specific $N$, although this varies with the number of particles. This implies that the average magnetisation is zero. When a small magnetic field $h$ is applied, the ground state is no longer shared by all states with the maximum $J$, but only by that for which $m=J$. Thus, the magnetisation reaches its maximum possible value and is maintained for any value $h>0$ and for any $q$ deformation parameter. In fact, the larger the deformation, the less energetic the previous state is.

\subsection{\texorpdfstring{$q$}{q}-deformation at the thermodynamic limit}
This subsection is devoted to characterizing the ferromagnetic phase transition in the thermodynamic limit of the $q$-deformed framework. Our primary focus is to discuss the evolution of the Curie transition and the magnetic and thermal properties in this limit. This analysis is crucial, as the thermodynamic limit provides the appropriate regime to observe the emergence of collective macroscopic behavior and to understand how $q$-deformation modifies the classic phase transition signatures. As stated in \cite{Ballesteros:2025cia}, the deformed energy eigenvalues are now given by $q$-numbers, which grow exponentially for large values of $\eta$. Therefore, the simplest form to assure the {extensivity} of the model for an arbitrary $N$ consists of defining the coupling constant of the deformed model as $I\rightarrow I/N$ (as in the undeformed KS model) and the deformation parameter as $\eta\rightarrow\eta/N$. {By extensivity, we mean that energy scales proportional to number of particles $N$}. Thus, in the following thermodynamic limits, we consider the $q$-KS model with the deformation parameter $q=e^{\eta/N}$.


\subsubsection{Thermodynamic properties}
\begin{figure}[H]
    \begin{minipage}{0.4\textwidth}
      (a)
      \includegraphics[scale=0.7]{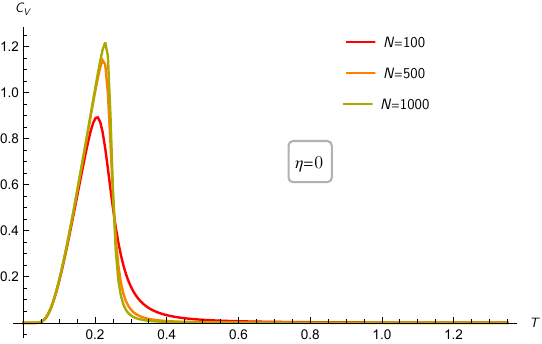}
    \end{minipage}
    \hspace{15mm}
    \begin{minipage}{0.4\textwidth}
      (b)
      \includegraphics[scale=0.7]{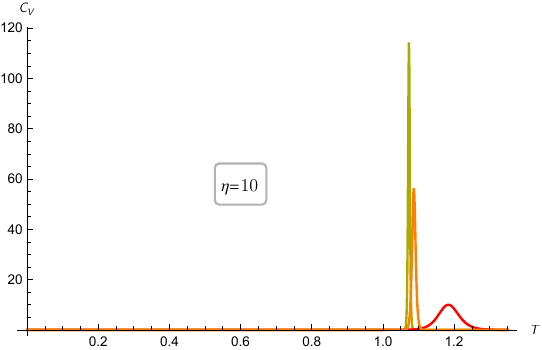}
    \end{minipage}
    \caption{Specific heat as a function of temperature for values of the number
    of particles $N=100$ (red), $500$ (orange), and $1000$ (yellow) for the
    interaction constant $I=1/N$ ferromagnetic case. The values of the parameter
    $q=e^{\eta/N}$ are: (a) $\eta=0$ and (b) {$\eta=10$}.}
    \label{cvlimitdeformed}
  \end{figure}

\begin{figure}[H]
    \begin{minipage}{0.4\textwidth}
      (a)
      \includegraphics[scale=0.7]{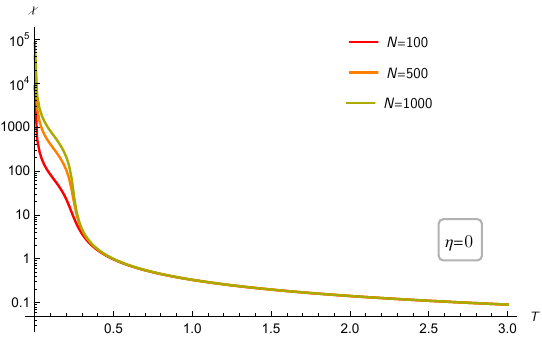}
    \end{minipage}
    \hspace{15mm}
    \begin{minipage}{0.4\textwidth}
      (b)
      \includegraphics[scale=0.7]{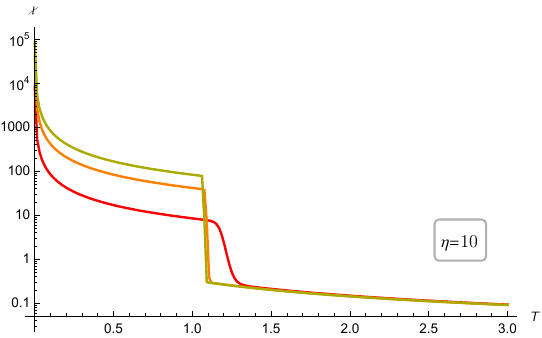}
    \end{minipage}
    \caption{Magnetic susceptibility as a function of temperature for values of
    the number of particles $N=100$ (red), $500$ (orange), and $1000$ (yellow)
    for the interaction constant $I=1/N$ ferromagnetic case. The values of the parameter
    $q=e^{\eta/N}$ are: (a) $\eta=0$ and (b) $\eta=10$.}
    \label{chilimitdeformed}
  \end{figure}

\begin{figure}[H]
    \begin{minipage}{0.4\textwidth}
      (a)
      \includegraphics[scale=0.7]{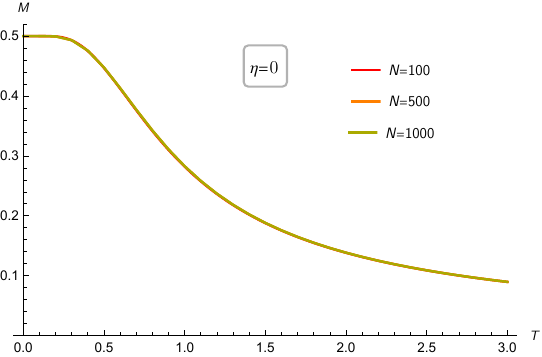}
    \end{minipage}
    \hspace{15mm}
    \begin{minipage}{0.4\textwidth}
      (b)
      \includegraphics[scale=0.7]{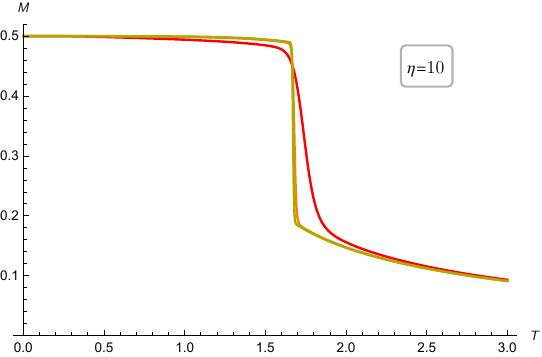}
    \end{minipage}
    \caption{Magnetisation as a function of temperature for values of the number
    of particles $N=100$ (red), $500$ (orange), and $1000$ (yellow) for the
    interaction constant $I=1/N$ and ${h\gamma=1}$ ferromagnetic case. The values of the
    parameter $q=e^{\eta/N}$ are: (a) $\eta=0$ and (b) $\eta=10$.}
    \label{mlimitdeformed}
  \end{figure}

\begin{figure}[H]
    \begin{minipage}{0.4\textwidth}
      (a)
      \includegraphics[scale=0.7]{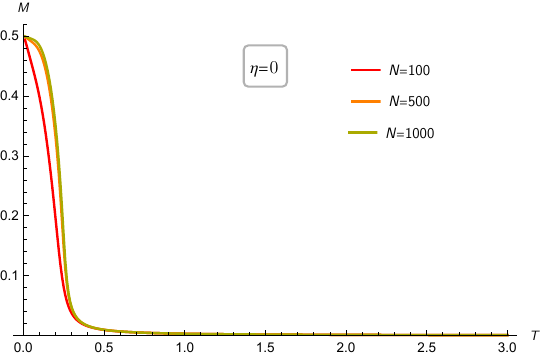}
    \end{minipage}
    \hspace{15mm}
    \begin{minipage}{0.4\textwidth}
      (b)
      \includegraphics[scale=0.7]{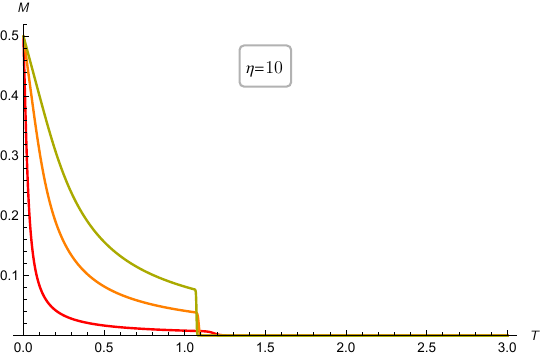}
    \end{minipage}
    \caption{Magnetisation as a function of temperature for values of the number
    of particles $N=100$ (red), $500$ (orange), and $1000$ (yellow) for the
    interaction constant $I=1/N$ and ${h\gamma=0}$ ferromagnetic case. The values
    of the parameter $q=e^{\eta/N}$ are: (a) $\eta=0$ and (b) $\eta=10$.}
    \label{mlimitedeformedh0}
  \end{figure}

{As in the previous subsections, all the thermodynamic quantities (except the Curie temperature) in this subsection are obtained by direct numerical evaluation of the partition function~\eqref{partition}.} In the undeformed case, the thermodynamic properties exhibit a behaviour in which the number of particles has little to no influence. In contrast, in the deformed case, it has been observed that the curves corresponding to different particle numbers are significantly more separated (as we will see in Figs.~\ref{cvlimitdeformed},\ref{chilimitdeformed}, \ref{mlimitdeformed}, and \ref{mlimitedeformedh0}). Moreover, this effect becomes more pronounced as the deformation increases.

The Curie temperature was obtained in \cite{Ballesteros:2025cia}. For small and intermediate deformation parameters ($\eta \in (0,8)$), it is approximated as (see \cite{Ballesteros:2025cia} for the derivation and details of these results)
  \begin{align}
    T_{C}(\eta)= & 0.25                                                                                  & \eta\in(0,3), \label{tc1} \\
    T_{C}(\eta)= & 0.25+0.022959\eta-0.023077\eta^{2}+0.007164\eta^{3}-0.000779\eta^{4}+0.000035\eta^{5} & \eta\in(3,8), \label{tc2}
  \end{align}
while for larger values of $\eta$ is
  \begin{equation}
    \lim_{N\rightarrow \infty}T_{C}(\eta)=\frac{e^{\eta/2}}{\eta^{2}\log (4)}. \label{tc3}
  \end{equation}
This temperature can be observed in the different thermodynamic magnitudes considered here, as we will see.

  For example, the peaks of the specific heat reach significantly higher values
  and occur at higher temperatures, although they also become narrower, as shown
  in Fig. \ref{cvlimitdeformed}. This behaviour arises because deformation
  broadens the energy distribution, reducing the influence of highly excited states
  and amplifying the action of the ground states within the considered temperature
  range.

{The value $\eta=10$ was specifically chosen to illustrate the sensitivity of the thermodynamic properties to the deformation parameter. This value is large enough to show a clear displacement of the phase transition compared to the undeformed case ($\eta=0$), yet small enough to ensure that the Curie transition remains within the observed temperature range ($T \in [0, 3]$).}

In terms of magnetic susceptibility, the introduction of deformation distorts the coincidence of the curves for different $N$ observed in the undeformed case. This coincidence occurs for values before and after the Curie temperature, and the deformation shifts this convergence to lower and higher temperatures. This effect is amplified as the deformation parameter increases. As expected, the Curie transition occurs at progressively higher temperatures as the deformation increases. It can be observed that the transition temperature values in the susceptibility coincide with the temperature of the peaks in the specific heat for the same values of deformation and particle number, as expected. Regarding its shape, it becomes progressively less smooth, with the curve appearing increasingly more vertical as the deformation increases.

In the undeformed case, all the magnetisation curves almost coincide, but they do not in the deformed case. Because the ground states remain the same in the deformed case (for a given $N$), a similar behaviour is observed at low temperatures for each deformation value. {Figures \ref{mlimitdeformed} (a) and (b)} take an external magnetic field of $h=1$, so we cannot see the Curie transition. Therefore, we reproduce these figures for a null external magnetic field.

We observe spontaneous magnetisation in the absence of an external magnetic field. As the temperature increases, the magnetisation gradually decreases until it reaches the Curie transition, at which a sharp drop occurs, ultimately reducing the magnetisation to zero.
{ 
\subsection{Finite-Size Scaling Analysis}
\label{sec:fss}

In order to determine the critical behavior of the system in the thermodynamic limit ($N \to \infty$), a systematic finite-size scaling analysis was performed for different values of the deformation parameter $\eta \in [1, 20]$. The numerical procedure relies on the exact evaluation of the canonical partition function for large system sizes, specifically $N \in \{10000, 30000, 60000, 100000\}$. For each size $N$ and parameter $\eta$, the specific heat $C_V(T)$ and the magnetic susceptibility $\chi(T)$ are computed to locate and analyze their corresponding peaks.

The extraction of the critical temperature and the scaling exponents is carried out through a multi-step statistical procedure. First, to optimize the localization of the maxima, the expressions of $T_C(\eta)$  given in Eqs.\eqref{tc1} and \eqref{tc2} are used as an analytical seed. Within a narrow temperature window around this seed, a fine mesh and parabolic interpolation are implemented to accurately extract the precise pseudo-critical peak temperature $T_{\max}(N)$ and the corresponding maximum peak heights. Subsequently, the critical temperature in the thermodynamic limit $T_c(\infty)$ is extrapolated through a linear regression of $T_{\max}(N)$ against $1/\sqrt{N}$. Finally, the scaling exponents $p(C_V)$ and $p(\chi)$ are extracted via a linear regression in a double-logarithmic scale of the maximum observable values against the system size $N$:
\begin{equation}
    \ln(\text{Observable}_{\max}) = p \ln(N) + B,
\end{equation}
where $p$ is the scaling exponent (the slope of the fit) and $B$ is a constant, implying that $\text{Observable}_{\max} \propto N^p$. The reliability and {accuracy} of these regressions are quantified through the coefficient of determination $R^2$.

The global results obtained from this systematic scaling analysis are summarized in Table~\ref{tab:fss_results}.

\begin{table}[H]
\centering
\small
\begin{tabular}{cccccc}
\hline
$\eta$ & $T_c(\infty)$ & $p(C_{V})$ & $R^2(C_{V})$ & $p(\chi)$ & $R^2(\chi)$ \\
\hline
1 & 0.2484 & 0.0259 & 0.9864 & 0.9999 & $\approx 1$ \\
2 & 0.2493 & 0.0522 & 0.9875 & 0.9999 & $\approx 1$ \\
3 & 0.2506 & 0.8556 & 0.9888 & 0.9999 & $\approx 1$ \\
4 & 0.2678 & 0.9941 & 0.9994 & 0.9999 & $\approx 1$ \\
5 & 0.3073 & 1.0001 & $\approx 1$    & 0.9999 & $\approx 1$ \\
6 & 0.3703 & 0.9473 & 0.9960 & 0.9999 & $\approx 1$ \\
7 & 0.4629 & 0.9808 & 0.9997 & 0.9999 & $\approx 1$ \\
8 & 0.5958 & 0.8876 & 0.9895 & 0.9999 & $\approx 1$ \\
9 & 0.7855 & 0.9962 & 0.9999 & 0.9999 & $\approx 1$ \\
10 & 1.0570 & 0.9808 & 0.9995 & 0.9999 & $\approx 1$ \\
11 & 1.4469 & 0.9611 & 0.9996 & 0.9999 & $\approx 1$ \\
12 & 2.0103 & 0.9158 & 0.9913 & 0.9999 & $\approx 1$ \\
13 & 2.8291 & 0.9807 & 0.9999 & 0.9999 & $\approx 1$ \\
14 & 4.0262 & 0.9914 & 0.9992 & 0.9999 & $\approx 1$ \\
15 & 5.7863 & 0.9883 & 0.9995 & 0.9999 & $\approx 1$ \\
16 & 8.3880 & 0.9228 & 0.9928 & 0.9999 & $\approx 1$ \\
17 & 12.2532 & 0.9959 & $\approx 1$   & 0.9999 & $\approx 1$ \\
18 & 18.0223 & 0.9995 & $\approx 1$   & 0.9999 & $\approx 1$ \\
19 & 26.6703 & 0.9913 & 0.9999 & 0.9999 & $\approx 1$ \\
20 & 39.6862 & 0.9846 & 0.9997 & 0.9999 & $\approx 1$ \\
\hline
\end{tabular}
{\caption{Global results of the finite-size scaling analysis for different values of the deformation parameter $\eta$ and for the ferromagnetic case.}
\label{tab:fss_results}}
\end{table}

The quantitative analysis of the data reveals distinct scaling regimes in the critical behavior. The magnetic susceptibility shows an extremely robust behavior across the entire parameter space, yielding \(p(\chi) \approx 1.0\) with an excellent fit quality (\(R^2 \approx 1.0\)). In contrast, the specific heat exponent \(p(C_{V})\) exhibits a clear crossover: for \(\eta = 1, 2\), the peak growth is very weak, while for \(\eta \ge 4\), the exponent firmly converges to \(p(C_{V}) \approx 1.0\) (\(R^2 \to 1\)). Consequently, in this strongly deformed regime, the specific heat peaks also scale linearly with \(N\). The crossover observed around \(\eta \approx 4\), which triggers this extensive divergence, is directly related to a structural change in the energy distribution. While for small deformation the Boltzmann weights exhibit a single peak, corresponding to a standard thermodynamic behavior, increasing \(\eta\) leads to a flattening of the distribution followed by the emergence of a bimodal structure \cite{Ballesteros:2025cia}.
}

\section{The antiferromagnetic case}
\label{sec:kskanti} 
We now move to the antiferromagnetic case for both undeformed and deformed Hamiltonians, and for a small number of spins and its thermodynamic limit.

\subsection{Small number of spins}
\begin{figure}[H]
    \centering
    \includegraphics[scale=0.9]{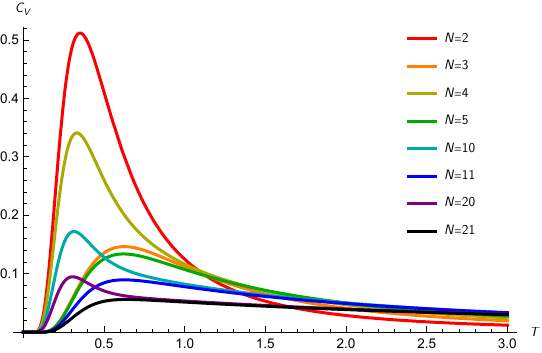}
    \caption{Specific heat as a function of temperature for values of the number
    of particles $N=2$ (red), $3$ (orange), $4$ (yellow), $5$ (green), $10$ (cyan),
    $11$ (blue), $20$ (purple), and $21$ (black) for the antiferromagnetic case.}
    \label{kskj1/2cvp}
  \end{figure}

We start by reproducing the results of~\cite{al1998exact} and extend them by computing the magnetisation and studying the corresponding behaviour at zero temperature. {Using Eqs.~\eqref{eq:thermody} and the numerical evaluation of the exact partition function, the graphs in that article corresponding to the specific heat can be easily reproduced}, as shown in Fig.~\ref{kskj1/2cvp}.
We observe a completely different behaviour if $N$ is odd or even. In the antiferromagnetic case, the ground states depend on the parity of the number of particles: for an even number of particles, they correspond to $J=0,1$, whereas for an odd number, they correspond to $J=1/2,3/2$. For $N$ even, the maximum is shifted to the left as $N$ increases. The position of this maximum depends mostly on the distance between the two lowest energy levels and, to a lesser extent, on the rate of degeneration of the two levels. This difference is $I$ for $N$ even and $3I/2$ for $N$ odd, which quantifies the distance between the corresponding maxima. At high temperatures, a behaviour $1/T^{2}$ is observed as long as the number of particles is small (see the analytical expression of Eq.~\eqref{eq:cvp}) because the amplitude of the energy levels is proportional to $N^{2}$; therefore, the temperature range considered here does not allow the system to reach the highest energy levels when $N$ increases \cite{al1998exact}.

\begin{figure}[H]
    \centering
    \includegraphics[scale=0.9]{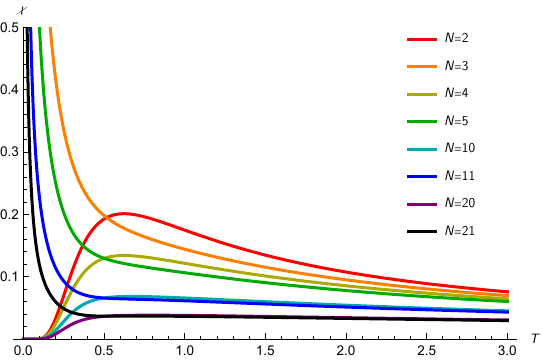}
    \caption{Magnetic susceptibility as a function of temperature for values of
    the number of particles $N=2$ (red), $3$ (orange), $4$ (yellow), $5$ (green),
    $10$ (cyan), $11$ (blue), $20$ (purple), and $21$ (black) for the antiferromagnetic
    case.}
    \label{kskj1/2xp}
  \end{figure}

For the magnetic susceptibility, we observe in Fig.~\ref{kskj1/2xp} also a significant difference between even and odd $N$. This behaviour can be understood as follows. For odd $N$, the lowest energy level always splits into two in the presence of the magnetic field, whereas for even $N$ it does not. In the odd-$N$ case, one can have configurations with as many spins up as down except for a single spin, so that the local field produced by the remaining spins on that spin vanishes (frustration) and only the external field acts on it, as discussed in Ref.~\cite{al1998exact}. Moreover, for $N$ odd, we observe a behaviour of $1/(NT)$, while in both cases, the behaviour decreases as $1/T$ at high temperatures (see the analytical expression of Eq.~\eqref{eq:chi_para}).

\begin{figure}[H]
    \centering
    \includegraphics[scale=0.9]{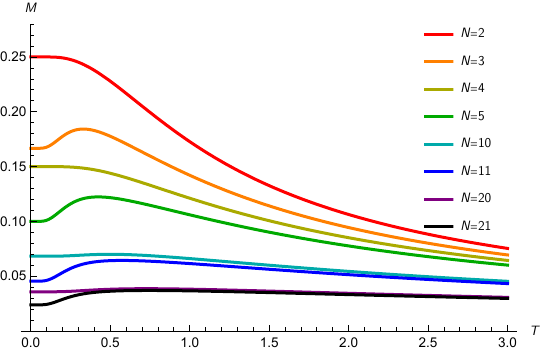}
    \caption{Magnetisation as a function of temperature for values of the number
    of particles $N=2$ (red), $3$ (orange), $4$ (yellow), $5$ (green), $10$ (cyan),
    $11$ (blue), $20$ (purple), and $21$ (black), with ${h\gamma=1}$ for
    antiferromagnetic case.}
    \label{kskj1/2mp}
  \end{figure}

The magnetisation is shown in Fig.~\ref{kskj1/2mp}, for an external magnetic field (in {$h\gamma=1$} units). In this scenario, we notice a slight bump at low temperatures when $N$ is odd, whereas no such bump appears when $N$ is even. Moreover, in both cases, it goes to zero for increasing $N$ and $T$. As before, the main contribution to the magnetisation is from the two lower bands.

The fact that at $T\to 0$ there is a maximum for $N$ even while there is a small bump for greater values of $T$ for $N$ odd can be understood from the following argument. For simplicity, we consider the cases $N=2$ and $N=3$. In the first case, the minimum energy is obtained for two different configurations: one spin up and one spin down, located in the $j=m=0$ state (with energy $-3/4$), and both spins up, on the $j=m=1$ (with energy $1/4-h$). Therefore, the magnetisation at zero temperature is maximum, corresponding to the average of the last two configurations. When the temperature increases, the other configurations come into play; therefore, the magnetisation decreases, going to zero for a sufficiently high temperature (the average of the magnetisation for all the possible configurations is zero).

For the $N=3$ case, the minimum energy configuration corresponds to the (doubly degenerate) $J=m=1/2$ state (with energy $-3/4-h/2$), with two spins up and one spin down. The first excited energy state corresponds to the three-spin-up configuration with $J=m=3/2$ (with energy $3/4-3h/2$). {Therefore, when the temperature increases slightly, the magnetisation increases because the first excited state ($m=3/2$) becomes thermally populated, increasing its contribution to the statistical average.} This explains the small bump in the magnetisation for $N$ odd. As in the $N$ even case, the magnetisation for high temperatures goes to zero because the other configurations are also possible.

Following this argument, one can understand the behaviour for $N$ even or odd for a larger number of spins.

The behaviour of these three quantities, and in particular the fact that the lower energy levels play a prominent role, can be understood from the energy spectrum~\eqref{eq:energy}. One can notice that the most probable levels, those with the lowest energy, are those with the smallest $J$.\footnote{It is important to note that states are defined just by angular momentum $J$ when an external magnetic field is not applied. This holds for the specific heat and magnetic susceptibility, but not for magnetisation. The states are described by the angular momentum $J$ and magnetic angular momentum $m$. The approximation we consider {in Eqs. \ref{eq:Part}} takes only the $J$ values (no the $m$ ones) of the two most probable levels, so we always refer to it as the approximation of the most probable levels, since more than two states (due to the $m$ degeneration) are indeed considered.} Therefore, for even $N$ the greatest contribution is given by the levels with $J=0,1$, and for odd $N$ the ones with $J=1/2,3/2$. For increasing $N$, the dominant term in the energy \eqref{eq:energy} becomes the one proportional to $N$. This allows us to consider an approximation of the partition function, for which the only terms that are considered are these, finding
  \begin{align}
    Z^{e}{\approx} & e^{-\frac{8 h-3 N+8}{8 T}}\left(\binom{N}{\frac{N}{2}}e^{\frac{h+1}{T}}-\binom{N}{\frac{N}{2}-2}\left(e^{h/T}+e^{\frac{2 h}{T}}+1\right)+\binom{N}{\frac{N}{2}-1}\left(e^{h/T}+e^{\frac{2 h}{T}}-e^{\frac{h+1}{T}}+1\right)\right),\notag                                              \\
    Z^{o}{\approx} & \left(e^{h/T}+1\right) \left(-e^{\frac{3 (-4 h+N-5)}{8 T}}\right) \left(\binom{N}{\frac{N-5}{2}}\left(e^{\frac{2 h}{T}}+1\right)+\binom{N}{\frac{N-3}{2}}\left(-e^{\frac{2 h}{T}}+e^{\frac{2 h+3}{2 T}}-1\right)-\binom{N}{\frac{N-1}{2}}e^{\frac{2 h+3}{2 T}}\right). \label{eq:Part}
  \end{align}
{The choice of taking the ground states is motivated by the fact that, at low temperatures, the Boltzmann weights are concentrated in these sectors due to the energy gap introduced by the interaction $I$. However, it is important to note that as $N \to \infty$, the number of states with low $J$ increases, and the gap effectively scales with the $I/N$ coupling. While this approximation captures the low-temperature scaling, its accuracy diminishes as $T$ increases where higher $J$ sectors begin to contribute to the total partition function.}

Therefore, we can obtain analytical expressions. In particular, we obtain the following for the specific heat:
  \begin{equation}
    C_{V}^{e} {\approx}\frac{9 (N+4) e^{1/T}}{\left(9 N T+(N+4) e^{1/T}T\right)^{2}},\qquad
    C_{V}^{o}{\approx}\frac{9 (N-1) (N+5) e^{\frac{3}{2 T}}}{N T^{2}\left((N+5) e^{\frac{3}{2
    T}}+4 (N-1)\right)^{2}}, \label{eq:cvp}
  \end{equation}
where the superscripts $e$ and $o$ correspond to the even and odd $N$ cases, respectively. The magnetic susceptibility reads
  \begin{equation}
    \chi^{e}{\approx}\frac{6}{9 N T+(N+4) e^{1/T}T},\qquad \chi^{o}{\approx}\frac{(N+5)
    e^{\frac{3}{2 T}}+20 (N-1)}{4 N T \left((N+5) e^{\frac{3}{2 T}}+4 (N-1)\right)}
    , \label{eq:chi_para}
  \end{equation}
and finally, the magnetisation is
  \begin{equation}
    M^{e}{\approx}\frac{3 \left(e^{2/T}-1\right)}{3 N e^{1/T}+4 (N+1) e^{2/T}+3 N},\qquad
    M^{o}{\approx}\frac{(N-1) \left(e^{1/T}-1\right) \left(e^{1/T}\left(3 e^{1/T}+4\right)+3\right)+(N+5)
    e^{3/T}\sinh \left(\frac{1}{2 T}\right)}{N \left(e^{1/T}+1\right) \left(2 (N-1)
    e^{2/T}+(N+5) e^{\frac{5}{2 T}}+2 (N-1)\right)}. \label{eq:m_para}
  \end{equation}

\begin{figure}[H]
    \begin{minipage}{0.4\textwidth}
      (a)
      \includegraphics[scale=0.7]{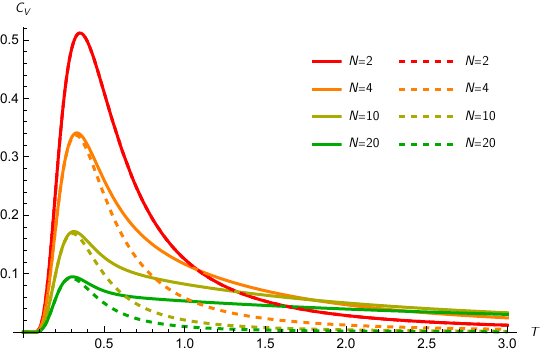}
    \end{minipage}
    \hspace{15mm}
    \begin{minipage}{0.4\textwidth}
      (b)
      \includegraphics[scale=0.7]{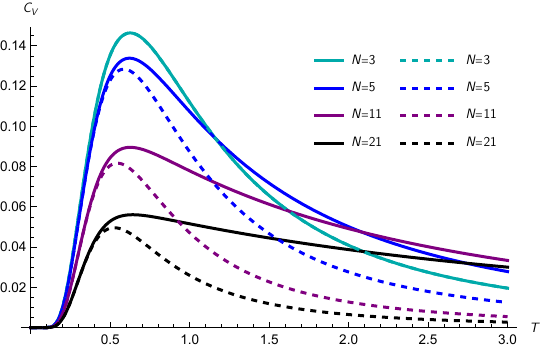}
    \end{minipage}
    \caption{Comparison of the exact specific heat (continuous line) with the
    most probable levels approximation (dashed line) as a function of
    temperature for values of (a) the even number of particles $N=2$ (red), $4$ (orange),
    $10$ (yellow) and $20$ (green) and (b) the odd number of particles $N=3$ (cyan),
    $5$ (blue), $11$ (purple), and $21$ (black) for the antiferromagnetic case.}
    \label{qkskj1/2comparisoncv}
  \end{figure}

\begin{figure}[H]
    \begin{minipage}{0.4\textwidth}
      (a)
      \includegraphics[scale=0.7]{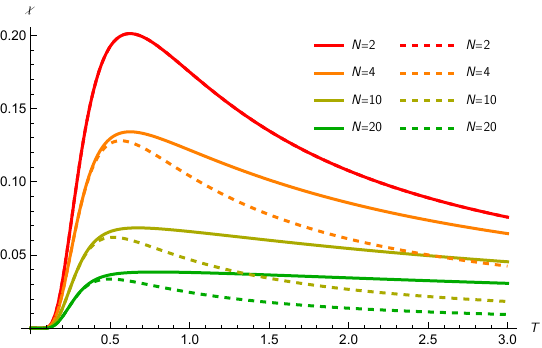}
    \end{minipage}
    \hspace{15mm}
    \begin{minipage}{0.4\textwidth}
      (b)
      \includegraphics[scale=0.7]{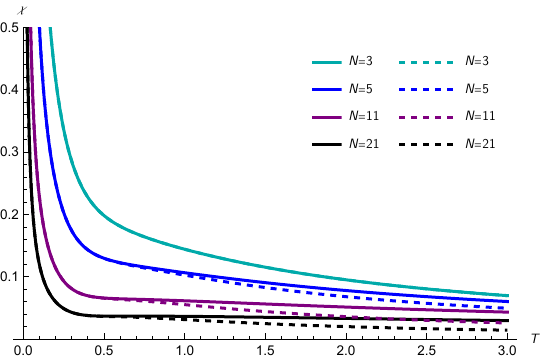}
    \end{minipage}
    \caption{Comparison of the exact magnetic susceptibility (continuous line)
    with the most probable levels approximation (dashed line) as a function of
    temperature for values of (a) the even number of particle $N=2$ (red), $4$ (orange),
    $10$ (yellow) and $20$ (green) and (b) the odd number of particles $N=3$ (cyan),
    $5$ (blue), $11$ (purple), and $21$ (black) for the antiferromagnetic case.}
    \label{qkskj1/2comparisonchi}
  \end{figure}

\begin{figure}[H]
    \begin{minipage}{0.4\textwidth}
      (a)
      \includegraphics[scale=0.7]{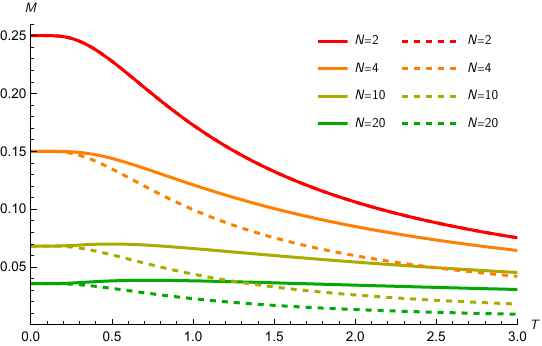}
    \end{minipage}
    \hspace{15mm}
    \begin{minipage}{0.4\textwidth}
      (b)
      \includegraphics[scale=0.7]{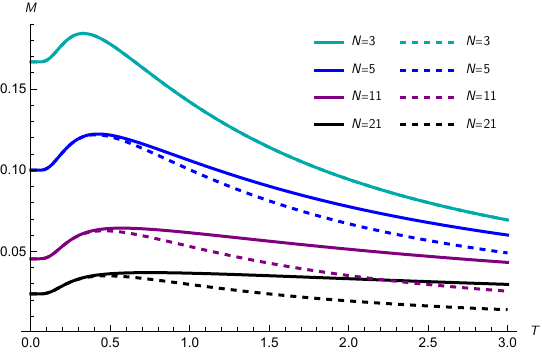}
    \end{minipage}
    \caption{Comparison of the exact magnetisation (continuous line) with the
    most probable levels approximation (dashed line) as a function of
    temperature for values of (a) the even number of particles $N=2$ (red), $4$ (orange),
    $10$ (yellow) and $20$ (green) and (b) the odd number of particles $N=3$ (cyan),
    $5$ (blue), $11$ (purple), and $21$ (black) for the ${h\gamma=1}$ antiferromagnetic
    case.}
    \label{qkskj1/2comparisonm}
  \end{figure}

Before proceeding, we compare the approximation carried out by considering the two most probable levels and the results obtained when the entire partition function is considered. To determine the {accuracy} of this approximation, the three thermodynamic properties considered above have been plotted, distinguishing between $N$ even and odd cases. This is shown in Figs. \ref{qkskj1/2comparisoncv}-\ref{qkskj1/2comparisonm}.

We can see that the approximation is really good for small temperature values and that it worsens as this magnitude increases, as well as when the number of particles considered is larger. Therefore, as discussed in the following, this approximation allows us to describe the behaviour of thermodynamic quantities at low temperatures.

\begin{figure}[H]
    \begin{minipage}{0.4\textwidth}
      (a)
      \includegraphics[scale=0.7]{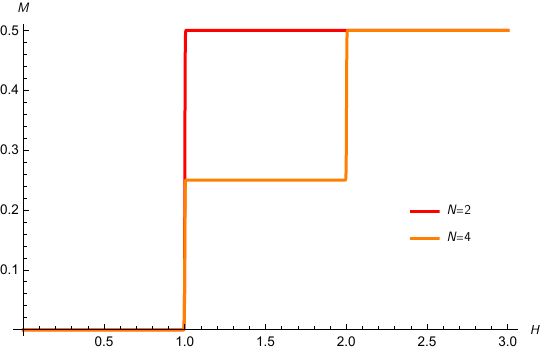}
    \end{minipage}
    \hspace{15mm}
    \begin{minipage}{0.4\textwidth}
      (b)
      \includegraphics[scale=0.7]{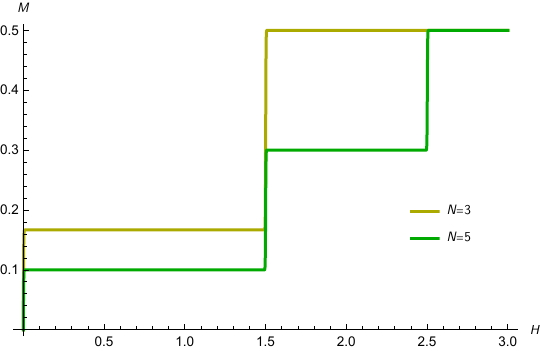}
    \end{minipage}
    \caption{Magnetisation as a function of magnetic field for (a) even values
    of the number of particles $N=2$ (red) and $4$ (orange) and (b) odd values
    $N=3$ (yellow), and $5$ (green) for temperature $T=0^{+}$, for the
    antiferromagnetic case.}
    \label{qkskj1/2mh}
  \end{figure}

We return to the discussion of the thermodynamic quantities from the approximated analytical expressions. It is clear from~\eqref{eq:chi_para} that the magnetic susceptibility vanishes as $T$ approaches zero for $N$ even, since the dominant term is $1/(T e^{1/T})$, while for $N$ odd it goes to infinity, going as $1/(4NT)$. Moreover, we can understand the magnetisation as a function of $T$, particularly for $T\to 0$. From~\eqref{eq:m_para}, we find that in this limit, the magnetisation for $N$ even is $3/(4(1+N))$, while for $N$ odd is $1/(2N)$. Moreover, because the magnetic susceptibility diverges as $T$ approaches zero, there are some ``jumps'' of the magnetisation at certain magnetic field values for zero temperature, as shown in Fig.~\ref{qkskj1/2mh}. These can be obtained from the energy spectrum of Eq.~\eqref{eq:energy}.

We first consider the even case. From Eq.~\eqref{eq:energy}, it is easy to observe that the lowest energy state (and therefore the one relevant for zero temperature) in the absence of an external magnetic field will be $J=m=0$. In this case, the magnetisation is zero because $m=0$ is the only possible quantum number. When the external magnetic field increases, there is a value of $h= {h}^{*}_{e}(1)$ for which the lowest-energy state becomes that with $J=m=1$. Therefore, a jump occurs for that value of the magnetic field. If the external magnetic field is increased again, up to ${h}^{*}_{e}(2)$, the lowest energy state will again be the one with one unit more of angular momentum, that is, $J=m=2$. This process is repeated, and for a sufficiently high external magnetic field, the lowest energy state is the one with $J=m=N/2$.

For $N$ odd, it is important to note that there is a jump at $h=h^{*}_{o}(1)=0$. Without an external magnetic field, the lowest energy levels are those with the lowest angular momentum ($J=1/2$); therefore, both states have the same energy. The total magnetisation of the system is zero because there is a cancellation between the two possible values of $m$. With a nonzero external magnetic field, the lowest-energy state is $J=m=1/2$, and thus, a jump occurs. When the external field is increased, the same behaviour as that in the even case is observed.

The number of jumps in the magnetisation that occur depends on the number of particles for both $N$ even and odd cases. In fact, there are $N/2$ or $(N+1)/2$ jumps, depending on whether $N$ is even or odd, respectively. The spacing between transitions is one unit of the magnetic field (as can be seen from~\eqref{eq:energy}), except for the spacing between the first and second jumps in the $N$ odd case, which is $h^{*}_{o}(2)-h^{*}_{o}(1)=3/2$ (remember that there is a phase transition in $h=h^{*}_{o}(1)=0$). The values of the external magnetic field at which the jumps take place are given by the following equations
  \begin{align}
    h^{*}_{o}(1)=0,\qquad h^{*}_{o}\left(\frac{n_{o}+1}{2}\right)=\frac{n_{o}}{2}, \qquad & \text{with}\qquad n_{o}=3,5,\dots , N, \label{kskj1/2msdo} \\
    h^{*}_{e}\left(\frac{n_{e}}{2}\right)=\frac{n_{e}}{2}, \qquad                         & \text{with}\qquad n_{e}=2,4,\dots, N. \label{kskj1/2msde}
  \end{align}
Fig.~\ref{qkskj1/2mh} shows this magnetisation behaviour as a function of the external magnetic field for $N=2,3,4,5$.
{The statements regarding the number of jumps and the general forms in Eqs.~\eqref{kskj1/2msdo} and~\eqref{kskj1/2msde} are based on an analytical computation (by observing the ground energy level as a function of the external magnetic field), and confirmed with numerical simulations across a wider range of system sizes than those explicitly plotted in Fig.~\ref{qkskj1/2mh}. We chose a lower value of $N$ in the figure specifically to ensure that the individual jumps and the structure of the steps remained clearly visible and distinguishable for the reader. These jumps correspond to the crossings of the energy levels in the spectrum.}

\subsection{Thermodynamic limit}
Following this analysis, the behaviours of the properties described above were studied in the thermodynamic limit {by numerical evaluation of the exact partition function}. In our case, the particle numbers $N=100, 1 01, 500, 501, 1000$ and $1001$ have been chosen, which are sufficiently large to study the thermodynamic limit {as we will see below}.

\begin{figure}[H]
    \centering
    \includegraphics[scale=0.9]{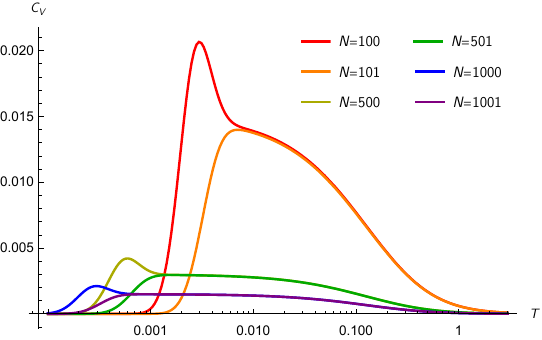}
    \caption{Specific heat as a function of temperature for values of the number
    of particles $N=100$ (red), $101$ (orange), $500$ (yellow), $501$ (green),
    $1000$ (blue), and $1001$ (purple), for the antiferromagnetic case.}
    \label{kskcvlt}
\end{figure}

In Fig.~\ref{kskcvlt}, we observe that, for a sufficiently high temperature, the behaviour of the specific heat for contiguous $N$ tends to be the same, regardless of whether $N$ is odd or even. We saw in Fig.~\ref{kskj1/2cvp} that saturation occurs after the maximum \cite{al1998exact}. This effect is maintained and scaled for high $N$, so the thermodynamic limit curves are lowered for increasing $N$.
  As seen in \cite{al1998exact}, the specific heat tends to zero in the
  thermodynamic limit, and there is still a maximum, corresponding to the two
  lowest energy levels, even for $N$ large. The behaviour decay $1/T^{2}$ is not
  observed, at least in the temperature range considered. For $N$ high enough, the
  energy distribution of the energy levels is compressed and becomes proportional
  to $N$ (as stated in \cite{al1998exact}), so that the temperature range
  considered here ($T\in(0,3)$) is sufficient to reach most of the highest
  energy levels. Thus, the system can hardly absorb more energy outside the
  considered temperature range.

  It is important to note that the most probable levels for the specific heat in
  the thermodynamic limit are still those with $J=0,1$ for $N$ even and
  $J=1/2, 3 /2$ for $N$ odd. Taking the approach of just the two most probable levels,
  the analytic expressions of the specific heat in the thermodynamic limit are
  \begin{equation}
    C_{V}^{e,\text{approx}}{\approx}\frac{9 e^{\frac{1}{N T}}}{N^{3}T^{2}\left(e^{\frac{1}{N T}}+9 \right)^{2}}
    ,\qquad C_{V}^{o,\text{approx}}{\approx}\frac{9 e^{\frac{3}{2 N T}}}{N^{3}T^{2}\left(e^{\frac{3}{2
    N T}}+4\right)^{2}}.
  \end{equation}

\begin{figure}[H]
    \centering
    \includegraphics[scale=0.9]{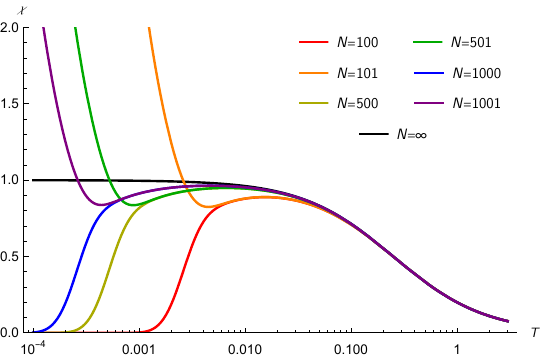}
    \caption{Magnetic susceptibility as a function of temperature for values of
    the number of particles $N=100$ (red), $101$ (orange), $500$ (yellow), $501$
    (green), $1000$ (blue), and $1001$ (purple), for the antiferromagnetic case.}
    \label{kskchilt}
  \end{figure}

In Fig.~\ref{kskchilt}, it can be seen how the magnetic susceptibility converges to the thermodynamic limit (discussed in~\cite{al1998exact}), defined by the line at which every considered case tends. We can distinguish notably the $N$ odd behaviour, which comes from above, from the $N$ even behaviour, which comes from below. In \cite{al1998exact}, the nonzero thermodynamic limit of the magnetic susceptibility is analytically obtained. It is shown that the magnetic susceptibility behaviour of $N\to\infty$ can be expressed as $\chi=1/(4T+1)$.

The change in coupling $I \to I/N$ does not imply changes in the most probable levels of the magnetic susceptibility once a magnetic field is applied, as this thermodynamic property is evaluated for a null field. Consequently, the most probable levels are still $J=0,1$ for even parity of $N$, and $J=1/2,3/2$ for odd parity of $N$. Then, the analytic expressions of magnetic susceptibility in the thermodynamic limit, taking the two most probable levels approximation, is as follows:
  \begin{equation}
    \chi^{e,\text{approx}}{\approx}\frac{6}{NT(9+e^{\frac{1}{N T}})},\qquad \chi^{o,\text{approx}}{\approx}\frac{e^{\frac{3}{2
    N T}}+20}{4 N T \left( e^{\frac{3}{2 N T}}+4 \right)}.
  \end{equation}

\begin{figure}[H]
    \centering
    \includegraphics[scale=0.9]{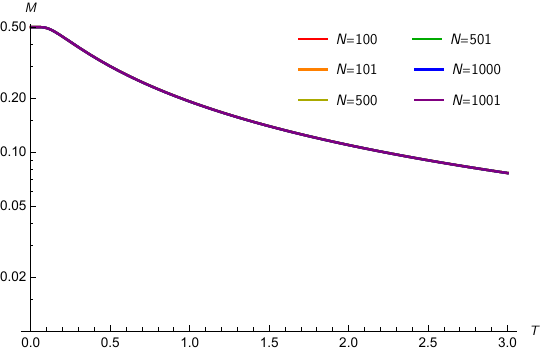}
    \caption{Magnetisation as a function of temperature for values of the number
    of particles $N=100$ (red), $101$ (orange), $500$ (yellow), $501$ (green),
    $1000$ (blue), and $1001$ (purple), for ${h\gamma=1}$ the antiferromagnetic
    case. They all are coincident.}
    \label{kskmlt}
  \end{figure}

In Fig.~\ref{kskmlt}, we present the magnetisation. We can see that all the $N$ curves coincide in the thermodynamic limit.
  
In the presence of an external magnetic field, the {ground} states are $J=m= N/2$ and $J=m=N/2-1$ {(which are not different from those in the ferromagnetic case)}. {The results for different sizes overlap almost perfectly} owing to the change in the coupling constant $I\rightarrow I/N$. Contrary to the previous quantities, the most probable states change in the magnetisation for the thermodynamic limit because an external magnetic field is applied. Then, the magnetic part of expression (\ref{eq:energy}) becomes the dominant term. For a small number of spins, we know that the minimum energy configuration is given by $J=m=1/2$ in the odd $N$ case and is shared between the $J=m=0$ and $J=m=1$ states in the even $N$ case. Introducing the coupling change $I\rightarrow I/N$, the most probable state becomes $J=m=N/2$ for any parity of $N$. This state is formed by $N$ spins up and it dominates in the small temperature range. Thus, the magnetisation takes the value $M=0.5$ at $T=0$ and decreases once the next energy level is reached, because the maximum value of $m$ decreases as $J$ does. This uniform decrease persists for the same reasons as in the ferromagnetic case because the energy differences between states remain the same for different values of $N$. Using the approximation of the most probable levels ($J=N/2,N/2-1$), the following analytical expression is obtained:
  \begin{equation}
    M^{\text{approx}}{\approx\frac{F+G}{H}}, \label{eq:magltferro}
  \end{equation}
  where
  \begin{equation*}
    {F}=(N-2) (N-1) e^{\frac{3}{2 T}}-N (N-1) e^{\frac{5}{2 T}}-(N+2) e^{1/T}+N,
  \end{equation*}
  \begin{equation*}
    {G}=e^{\frac{2 N+1}{2 T}}\left((N-2) (N-1) \left(-e^{1/T}\right)-N e^{\frac{3}{2
    T}}+(N+2) e^{\frac{1}{2 T}}+N (N-1)\right),
  \end{equation*}
  \begin{equation*}
    {H}=2 N \left(e^{1/T}-1\right) \left(e^{\frac{1}{2 T}}\left((N-1) e^{1/T}
    -e^{N/T}\left(N+e^{\frac{1}{2 T}}-1\right)\right)+1\right).
  \end{equation*}
{This is an equation not very accurate at high temperatures, but it reasonably reproduces the magnetisation at low temperatures before the transition.}

{The choice of system sizes up to $N=1000$ and $1001$ is sufficiently large to capture the thermodynamic limit of the system. As observed in the figures, the curves for large $N$ perfectly overlap (within the resolution of the plots) and become independent of the system size for most of the temperature range (for $T> 0.1$). The visible discrepancies at the extremely low-temperature regime ($T \to 0$), particularly in the specific heat and magnetic susceptibility, do not indicate a lack of convergence. Rather, they reflect the intrinsic parity effect of the antiferromagnetic Kittel-Shore model, where systems with even and odd $N$ possess fundamentally different ground states. Conversely, when an external magnetic field is applied (Fig. \ref{kskmlt}), the field energy dominates over these parity differences, resulting in perfectly coincident magnetisation curves across the entire temperature spectrum. This confirms that the macroscopic behavior is accurately captured.}

In the previous graphs of the specific heat and magnetic susceptibility in the thermodynamic limit, it is important to note that the maxima are shifted to the left for increasing values of $N$. For large $N$, all the thermodynamic properties~\eqref{eq:cvp}–\eqref{eq:m_para} converge to the same value, independent of the value of $N$, up to a $1/N$ factor. However, the interaction coupling $I$ is replaced by $I/N$ in the thermodynamic limit, which allows the displacement of the maxima. The magnetisation alone has coincident behaviour because the magnetic part becomes extremely dominant. Therefore, the behaviour of these quantities for larger $N$ can be easily inferred from the presented plots.

\begin{figure}[H]
    \begin{minipage}{0.4\textwidth}
      (a)
      \includegraphics[scale=0.7]{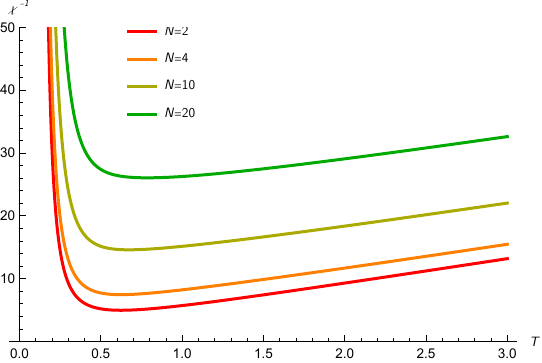}
    \end{minipage}
    \hspace{15mm}
    \begin{minipage}{0.4\textwidth}
      (b)
      \includegraphics[scale=0.7]{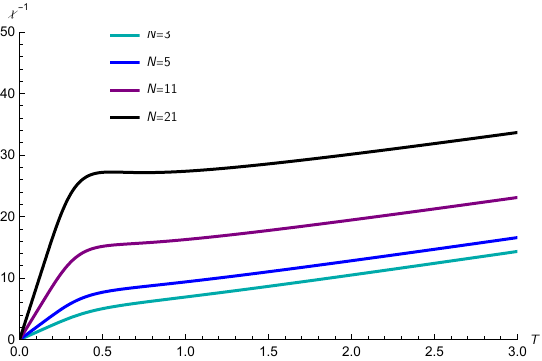}
    \end{minipage}
    \caption{Inverse of magnetic susceptibility as a function of temperature for
    values of (a) the even number of particles $N=2$ (red), $4$ (orange), $10$ (yellow)
    and $20$ (green) and (b) the odd number of particles $N=3$ (cyan), $5$ (blue),
    $11$ (purple), and $21$ (black), for the antiferromagnetic case.}
    \label{fig:curieanti}
  \end{figure}

The Curie temperature can now be obtained. Representing the inverse of the magnetic susceptibility, we find Fig.~\ref{fig:curieanti}.

From the the analytic expression obtained previously {(\ref{eq:chi_para})}, these are given by
  \begin{equation}
    \frac{1}{\chi^{e}}{\approx}\frac{9 N T+(N+4) e^{1/T}T}{6},\qquad \frac{1}{\chi^{o}}{\approx}\frac{4N T \left((N+5) e^{\frac{3}{2 T}}+4 (N-1)\right)}{(N+5) e^{\frac{3}{2 T}}+20(N-1)},
  \end{equation}

The even susceptibility does not diverge; therefore, it is not reasonable to calculate the Curie temperature in this case. In the odd case, the Curie temperature is $T_{C}=0$ and the Curie constant is $C=\frac{1}{4N}$. It is important to note that in the thermodynamic limit, this would simply be an extension of what has been observed here and does not involve any particularly interesting discussion.

\subsection{\texorpdfstring{$q$}{q}-deformation for small number of spins}
As discussed in the previous section, the main contribution to the thermodynamic quantities in the antiferromagnetic case is given by the two lowest-energy bands. In the $q$-deformed scenario, it is important to notice that $[1/2]_{q}[3/2]_{q}\to 1$ when $\eta (q)\to \infty$. Therefore, the term proportional to $N$ in Eq.~\eqref{eq:qenergy} is negligible for large values of $\eta$. As in the undeformed case, the main contributions are given by $J=1/2,3/2$ or $J=0,1$ for $N$ odd or even, respectively. Then, we find the following approximated partition functions:
  \begin{align}
    Z^{e}_{q} & {\approx}e^{-\frac{2 h-N \left[\frac{1}{2}\right]_{q}\left[\frac{3}{2}\right]_{q}+[2]_{q}+1}{2 T}}\left(\binom{N}{\frac{N}{2}}e^{\frac{2 h+[2]_{q}}{2 T}}+\binom{N}{\frac{N}{2}-1}\left(-e^{\frac{2 h+[2]_{q}}{2 T}}+e^{\frac{2 h+1}{2 T}}+e^{\frac{4 h+1}{2 T}}+e^{\frac{1}{2 T}}\right)-\right.\notag                                                                                                                                                       \\
              & \left.-e^{\frac{1}{2 T}}\binom{N}{\frac{N}{2}-2}\left(e^{h/T}+e^{\frac{2 h}{T}}+1\right)\right),\notag                                                                                                                                                                                                                                                                                                                                                \\
    Z^{o}_{q} & {\approx}\left(e^{h/T}+1\right) \left(\binom{N}{\frac{N-5}{2}}\left(e^{\frac{2 h}{T}}+1\right) e^{\frac{\left[\frac{1}{2}\right]_{q}\left[\frac{3}{2}\right]_{q}}{2 T}}-\binom{N}{\frac{N-3}{2}}\left(e^{\frac{4 h+\left[\frac{1}{2}\right]_{q}\left[\frac{3}{2}\right]_{q}}{2 T}}-e^{\frac{2 h+\left[\frac{3}{2}\right]_{q}\left[\frac{5}{2}\right]_{q}}{2 T}}+e^{\frac{\left[\frac{1}{2}\right]_{q}\left[\frac{3}{2}\right]_{q}}{2 T}}\right)-\right.\notag \\
              & \left.-\binom{N}{\frac{N-1}{2}}e^{\frac{2 h+\left[\frac{3}{2}\right]_{q}\left[\frac{5}{2}\right]_{q}}{2 T}}\right) \left(-e^{\left(-\frac{3 h-(N-1) \left[\frac{1}{2}\right]_{q}\left[\frac{3}{2}\right]_{q}+\left[\frac{3}{2}\right]_{q}\left[\frac{5}{2}\right]_{q}}{2 T}\right)}\right).
  \end{align}

Considering this argument, and as in the undeformed case, it was verified that the representations of the numerical results are faithfully reproduced with only the most probable levels in the antiferromagnetic case. Thus, we can obtain approximate analytic expressions for the specific heat as follows:
  \begin{align}
    C_{V,q}^{e} & {\approx}\frac{9 (N+4) (q+1)^{2}e^{\frac{q^{2}+1}{2 T \sqrt{q} (1-q)}}}{4 q \left(9 N T e^{\frac{q^{3/2}}{2 T(1-q)}}+(N+4) T e^{\frac{1}{2 T\sqrt{q}(1-q)}}\right)^{2}}=\notag                                                                                                                                                     \\
                & =C_{V}^{e}+\frac{9 (N+4) e^{1/T}\left((N+4) e^{1/T}(2 T-1)+9 (2 N T+N)\right)}{8 T^{3}\left((N+4) e^{1/T}+9 N\right)^{3}}(q-1)^{2}+O\left((q-1)^{3}\right),                                                                                                                                                                \\
    C_{V,q}^{o} & {\approx}\frac{(N-1) (N+5) \left(q^{2}+q+1\right)^{2}e^{\frac{(q+1) \left(q+\sqrt{q}+1\right)^2}{2 \left(\sqrt{q}+1\right)^2 q T}}}{N q^{2}T^{2}\left((N+5) e^{\left(\frac{\left(1-q^{3/2}\right) \left(1-q^{5/2}\right)}{2 (q-1)^2 q T}\right)}+4 (N-1) e^{\frac{q+\sqrt{q}+1}{2 \left(\sqrt{q}+1\right)^2 T}}\right)^{2}}=\notag \\
                & =C_{V}^{o}+\frac{3 (N-1) (N+5) e^{\frac{3}{2 T}}\left((N+5) e^{\frac{3}{2 T}}(4 T-3)+4 (N-1) (4 T+3)\right)}{2 N T^{3}\left((N+5) e^{\frac{3}{2 T}}+4 (N-1)\right)^{3}}(q-1)^{2}+O\left((q-1)^{3}\right).
  \end{align}
  The magnetic susceptibility is represented by
  \begin{align}
    \chi^{e}_{q} & {\approx}\frac{6}{T \left((N+4) e^{\frac{q+1}{2 \sqrt{q} T}}+9 N\right)}= \notag                                                                                                                                                                                                                                                                                 \\
                 & =\chi^{e}-\frac{3 (N+4) e^{\frac{1}{T}}}{4 T^{2}\left(N e^{\frac{1}{T}}+9 N+4 e^{\frac{1}{T}}\right)^{2}}(q-1)^{2}+O\left((q-1)^{3}\right),                                                                                                                                                                                                              \\
    \chi^{o}_{q} & {\approx}\frac{(N+5) e^{\left(\frac{\left(1-q^{3/2}\right) \left(1-q^{5/2}\right)}{2 (q-1)^2 q T}\right)}+20 (N-1) e^{\frac{q+\sqrt{q}+1}{2 \left(\sqrt{q}+1\right)^2 T}}}{4 N T \left((N+5) e^{\left(\frac{\left(1-q^{3/2}\right) \left(1-q^{5/2}\right)}{2 (q-1)^2 q T}\right)}+4 (N-1) e^{\frac{q+\sqrt{q}+1}{2 \left(\sqrt{q}+1\right)^2 T}}\right)}= \notag \\
                 & =\chi^{o}-\frac{2 \left(N^{2}+4 N-5\right) e^{\left.\frac{3}{2}\right/T}}{N T^{2}\left(N e^{\left.\frac{3}{2}\right/T}+4 N+5 e^{\left.\frac{3}{2}\right/T}-4\right)^{2}}(q-1)^{2}+O\left((q-1)^{3}\right).
  \end{align}
and the magnetisation is given by
  \begin{align}
    M^{e}_{q} & {\approx}\frac{3 e^{\frac{q^{3/2}}{2T(1-q)}}\left(e^{\frac{2}{T(1-q)}}-e^{\frac{2 q}{T(1-q)}}\right)}{3 N e^{\frac{4+q^{3/2}}{2T(1-q)}}+3 N e^{\frac{2(1+q) +q^{3/2}}{2 T(1-q)}}+3 N e^{\frac{q^{3/2}+4 q}{2 T(1-q)}}+(N+4) e^{\frac{2 (1+q)+\frac{1}{\sqrt{q}}}{2 T(1-q)}}}= \notag                                                                                                                                                                                                                                                                 \\
              & =M^{e}-\frac{3 (N+4) e^{2/T}\left(e^{2/T}-1\right)}{8 T \left(3 N e^{1/T}+4 (N+1) e^{2/T}+3 N\right)^{2}}(q-1)^{2}+O\left((q-1)^{3}\right),                                                                                                                                                                                                                                                                                                                                                                                                  \\
    M^{o}_{q} & {\approx}\frac{(N+5) \left(e^{\frac{1}{T}}-1\right) e^{\left(\frac{1}{T} +\frac{\left(1-q^{3/2}\right) \left(1-q^{5/2}\right)}{2 q (q-1)^2 T}\right)}+2 (N-1) e^{\frac{q+\sqrt{q}+1}{2 \left(\sqrt{q}+1\right)^2 T}}\left(-e^{\frac{1}{T}}+e^{\frac{2}{T}}+3 e^{\frac{3}{T}}-3\right)}{2 N \left(e^{\frac{1}{T}}+1\right) \left((N+5) e^{\left(\frac{1}{T} +\frac{\left(1-q^{3/2}\right) \left(1-q^{5/2}\right)}{2 q (q-1)^2 T}\right)}+2 (N-1) e^{\frac{q+\sqrt{q}+1}{2 \left(\sqrt{q}+1\right)^2 T}}\left(e^{\frac{2}{T}}+1\right)\right)}= \notag \\
              & =M^{o}-\frac{\left(N^{2}+4 N-5\right) e^{\left.\frac{5}{2}\right/T}\left(e^{\frac{1}{T}}-1\right) \left(e^{\frac{1}{T}}+1\right)}{N T \left(2 N e^{2/T}+N e^{\left.\frac{5}{2}\right/T}+2 N-2 e^{2/T}+5 e^{\left.\frac{5}{2}\right/T}-2\right)^{2}}(q-1)^{2}+O\left((q-1)^{3}\right).
  \end{align}

\begin{figure}[H]
    \begin{minipage}{0.4\textwidth}
      (a)
      \includegraphics[scale=0.7]{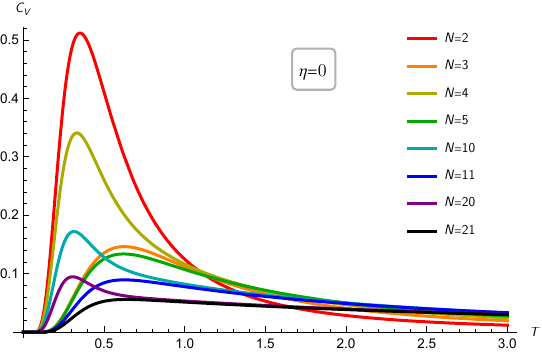}
    \end{minipage}
    \hspace{15mm}
    \begin{minipage}{0.4\textwidth}
      (b)
      \includegraphics[scale=0.7]{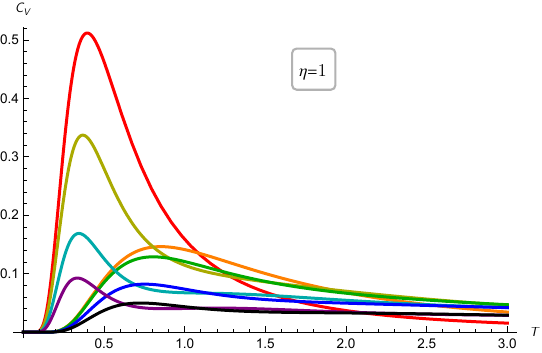}
    \end{minipage}
    \caption{Specific heat as a function of temperature for values of the number
    of particles $N=2$ (red), $3$ (orange), $4$ (yellow), $5$ (green), $10$ (cyan),
    $11$ (blue), $20$ (purple), and $21$ (black) and of the parameter (a) $\eta=0$
    and (b) $\eta=1$, for the antiferromagnetic case.}
    \label{qkskj1/2cvp}
  \end{figure}

We now represent the thermodynamic quantities {obtained from numerical computation}. We start by considering the specific heat, as shown in Fig.~\ref{qkskj1/2cvp} as a function of $\eta$.

{As expected, the undeformed case is fully recovered when $\eta=0$.} The values of the maxima for different $N$ are the same as in the undeformed case, and the different behaviours when $N$ is even or odd are preserved. Moreover, one can observe that the peak moves to the right as the parameter $\eta$ increases because the energy corresponding to the second lowest angular momentum becomes larger. This implies that the peaks for $\eta=0$ are placed farthest to the left.

\begin{figure}[H]
    \begin{minipage}{0.4\textwidth}
      (a)
      \includegraphics[scale=0.7]{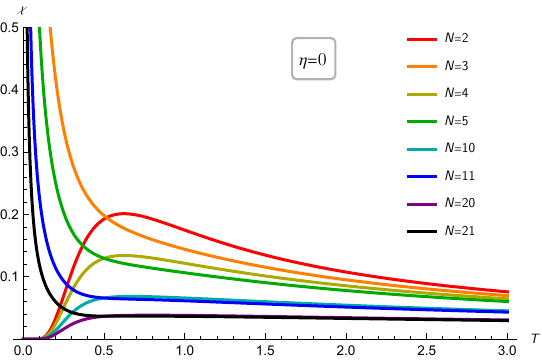}
    \end{minipage}
    \hspace{15mm}
    \begin{minipage}{0.4\textwidth}
      (b)
      \includegraphics[scale=0.7]{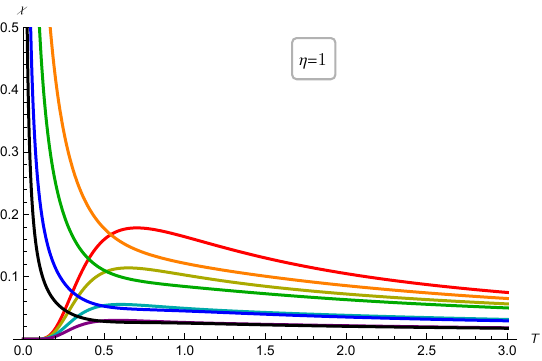}
    \end{minipage}
    \caption{Magnetic susceptibility as a function of temperature for values of
    the number of particles $N=2$ (red), $3$ (orange), $4$ (yellow), $5$ (green),
    $10$ (cyan), $11$ (blue), $20$ (purple), and $21$ (black) and of the parameter
    (a) $\eta=0$ and (b) $\eta=1$, for the antiferromagnetic
    case.}
    \label{qkskj1/2xp}
  \end{figure}

Regarding the magnetic susceptibility, Fig.~\ref{qkskj1/2xp} was obtained for different values of $\eta$. We observe that the maxima for $N$ even become smaller as $\eta$ increases, and for $N$ odd, the behaviour of $1/(NT)$ does not change as a function of $\eta$ for high values of $T$.
The maxima of $N$ even, which become smaller with increasing $\eta$, can be easily understood as follows. The energy of different states, except for the {ground} state, exponentially increases as the deformation parameter increases; therefore, they become less probable. Consequently, in the observed peaks corresponding to the transition from the ground state to the first excited state, the probability weight of the ground state is larger.

\begin{figure}[H]
    \begin{minipage}{0.4\textwidth}
      (a)
      \includegraphics[scale=0.7]{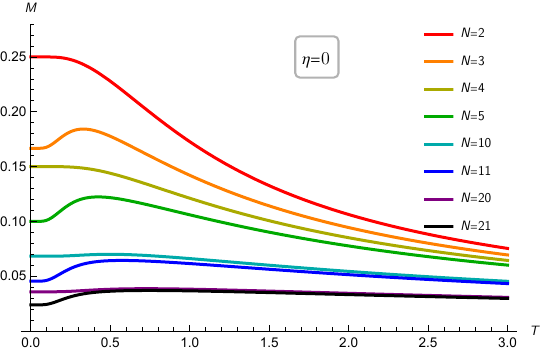}
    \end{minipage}
    \hspace{15mm}
    \begin{minipage}{0.4\textwidth}
      (b)
      \includegraphics[scale=0.7]{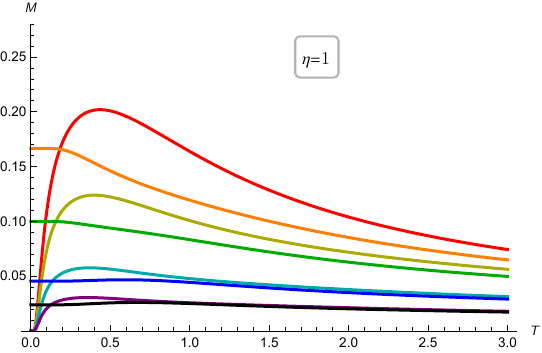}
    \end{minipage}
    \caption{Magnetisation as a function of temperature for values of the number
    of particles $N=2$ (red), $3$ (orange), $4$ (yellow), $5$ (green), $10$ (cyan),
    $11$ (blue), $20$ (purple), and $21$ (black) and of the parameter (a) $\eta=0$
    and (b) $\eta=1$ for the ${h\gamma=1}$
    antiferromagnetic case. }
    \label{qkskj1/2mp}
  \end{figure}

Finally, we compute the magnetisation in the presence of an external magnetic field in Fig.~\ref{qkskj1/2mp}. Here, we can observe an interesting feature: there is a completely different behaviour for $N$ odd or even when increasing $\eta$. For $N$ odd, the bump near zero temperature (commented in the previous section) disappears with the $\eta$ parameter large enough, the value at zero temperature being the same as in the undeformed case. For $N$ even, the maxima are shifted to the right and become smaller, and the magnetisation vanishes at zero temperature for $\eta\neq 0$.

We can now try to understand the behaviour of the magnetisation observed above for $h=1$ as a function of the deformation parameter $q$. As before, to understand this, we consider the $N=2$ and $N=3$ cases for a generic deformation parameter $q$. For the first case, one finds that the minimum energy state is not shared by the two configurations as in the undeformed scenario, but it is given by one spin up and one spin down with $J=m=0$ (and energy $-\left[1/2\right]_{q}\left[3/2\right]_{q}$). The next energy level is given by two spins up with $J=m=1$ (and energy $1/2([2]_{q}-\left[1/2\right]_{q}\left[3/2\right]_{q}-2)$). This last configuration {becomes thermally populated} when the temperature is sufficiently increased and, consequently, the magnetisation increases to the observed peak in every $N$ even case. When the temperature continues to increase, all possible configurations are averaged, and the magnetisation goes to zero.

As discussed above, for $N$ even, the magnetisation vanishes at zero temperature when a deformation is applied. In the undeformed model, a jump occurs at zero temperature and magnetic field $h=1$. However, in the deformed case, with $q\neq 1$, the same jump occurs for a magnetic field value larger than $h=1$ (see discussion below for the exact values of the magnetic field at which the jumps occur as a function of the deformation parameter). Therefore, the magnetisation for $N$ even is null because the first jump does not take place at $h=1$.

In the $N=3$ case, the minimum energy configuration is the same as in the undeformed case: two spins up and one spin down with $J=m=1/2$ (and energy $-\left[1/2\right]_{q}\left[3/2\right]_{q}-1/2$). This explains why the magnetisation in odd $N$ cases starts from the same value for both deformed and undeformed scenarios, independently of the value of $\eta$. However, the first excited state varies as a function of the deformation parameter $\eta$. When it belongs to the interval $\eta\in[0,0.9625)$, this state is the same as in the undeformed case: the three spins up with $J=m=3/2$ (and energy $1/2(\left[3/2\right]_{q}\left[5/2\right]_{q}-3\left[1/2\right]_{q}\left[3/2\right]_{q})-3/2$). Thus, it can be observed that for the representations of $\eta=0.1$ and $\eta=0.5$, there is a small peak when the temperature is slightly increased, as this state has greater magnetisation. However, when the deformation parameter is greater than the critical value, that is, for $\eta>0.9625$, the second lowest energy state is given by two spins down and one spin up with $J={-}m=1/2$ (with energy $-\left[1/2\right]_{q}\left[3/2\right]_{q}+1/2$). This configuration does not increase the magnetisation; therefore, for $\eta=1$ this quantity decreases smoothly from its initial value at a small temperature and then goes to zero as the temperature increases. The same explanation can be applied for all $N$ odd cases.

The value of the magnetisation at zero temperature for $N$ odd can also be understood by examining the magnetisation values at which the jumps occur. Since the first jump of the magnetisation for both deformed and undeformed cases takes place first at $h=0$, the magnetisation at zero temperature for $N$ odd has the same value as the undeformed case.

  \begin{figure}[H]
    \begin{minipage}{0.4\textwidth}
      (a)
      \includegraphics[scale=0.7]{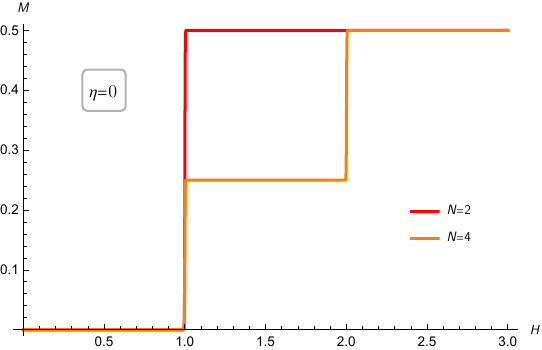}
    \end{minipage}
    \hspace{15mm}
    \begin{minipage}{0.4\textwidth}
      (b)
      \includegraphics[scale=0.7]{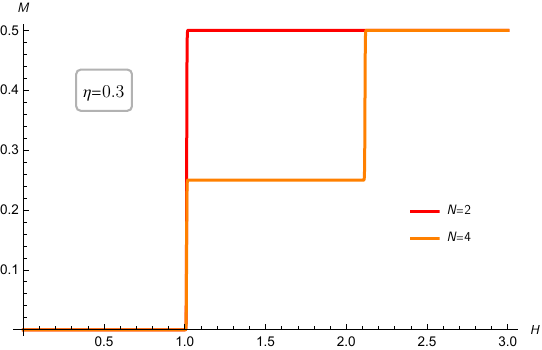}
    \end{minipage}
  \end{figure}
  \begin{figure}[H]
    \begin{minipage}{0.4\textwidth}
      (c)
      \includegraphics[scale=0.7]{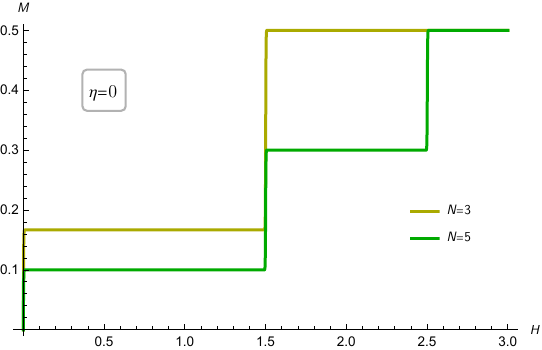}
    \end{minipage}
    \hspace{15mm}
    \begin{minipage}{0.4\textwidth}
      (d)
      \includegraphics[scale=0.7]{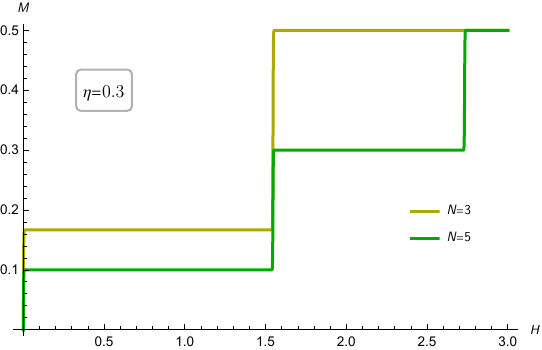}
    \end{minipage}
    \caption{Magnetisation as a function of magnetic field for values of the
    number of particles and of the parameter $\eta$ for temperature $T=0^{+}$, for
    the antiferromagnetic case. The representations (a) and (b) represent the
    even number of particles $N=2$ (red) and $4$ (orange) for $\eta=0$, $\eta=0.3$
    respectively, while (c) and (d) represent the odd number of particles $N=3$
    (yellow) and $5$ (green) for $\eta=0$, $\eta=0.3$ respectively.}
    \label{qkskj1/2transition}
  \end{figure}

The number of jumps in the magnetisation that occur depends exactly as in the undeformed case on the number of particles: there are $N/2$ or $(N+1)/2$ phase transitions depending on whether $N$ is even or odd, respectively. The difference with respect to the undeformed case is due to the spacing between transitions, because the energies for different angular momenta are not given by~\eqref{eq:energy} but~\eqref{eq:qenergy}. As before, by computing the difference between contiguous energy levels from~\eqref{eq:qenergy} it is easy to obtain the values of the external magnetic field at which the jumps take place
  \begin{align}
    h^{*}_{o}(1)=0,\qquad h^{*}_{o}\left(\frac{n_{o}+1}{2}\right)=\frac{[n_{o}]_{q}}{2}, \qquad & \text{with}\qquad n_{o}=3,5,\dots , N, \label{qkskj1/2msdo} \\
    h^{*}_{e}\left(\frac{n_{e}}{2}\right)=\frac{[n_{e}]_{q}}{2}, \qquad                         & \text{with}\qquad n_{e}=2,4,\dots, N. \label{qkskj1/2msde}
  \end{align}
  {As in the undeformed case, the statements regarding the number of jumps and Eqs.~\eqref{qkskj1/2msdo} and~\eqref{qkskj1/2msde} are based on analytical computations of the ground energy level as a function of the magnetic field and confirmed by numerical simulations for larger system sizes. A lower value of $N$ was chosen for Fig.~\ref{qkskj1/2transition} to ensure the individual jumps remain clearly distinguishable. These jumps correspond directly to the crossings of the energy levels in the $q$-deformed spectrum.} Of course, when taking the limit $q\to1$ one finds the same expressions~\eqref{kskj1/2msdo}--\eqref{kskj1/2msde} obtained in the previous section.

As can be seen in Fig.~\ref{qkskj1/2transition}, all jumps tend to move to the right, except for the first one (for zero magnetic field) in the $N$ odd case because, as commented above, the minimum energy configuration is shared by two states whose average magnetisation is zero. When a magnetic field is applied, the state with the highest $m$ becomes the one with the minimum energy; thus, a jump occurs at ${h}^{*}_{o}(1)=0$.

\subsection{\texorpdfstring{$q$}{q}-deformation of the thermodynamic limit}
\begin{figure}[H]
    \begin{minipage}{0.4\textwidth}
      (a)
      \includegraphics[scale=0.7]{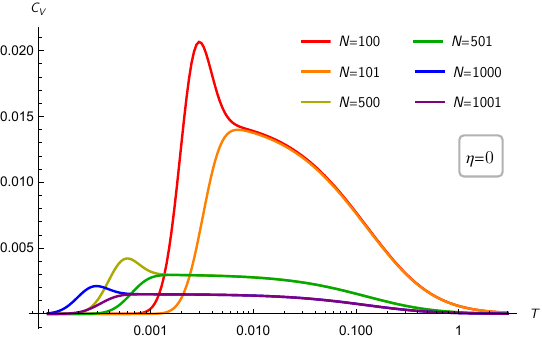}
    \end{minipage}
    \hspace{15mm}
    \begin{minipage}{0.4\textwidth}
      (b)
      \includegraphics[scale=0.7]{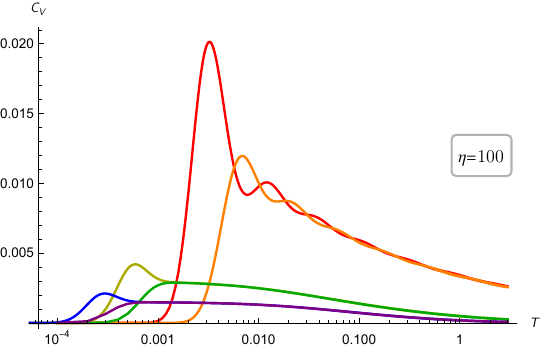}
    \end{minipage}
    \caption{Specific heat as a function of temperature for values of the number
    of particles $N=100$ (red), $101$ (orange), $N=500$ (yellow), $501$ (green),
    $N=1000$ (blue), $1001$ (purple) for the antiferromagnetic case. The values
    of the parameter $q=e^{\eta/N}$ are: (a) $\eta=0$ and (b) $\eta=100$.}
    \label{qkskj1/2transitioncv}
  \end{figure}

\begin{figure}[H]
    \begin{minipage}{0.4\textwidth}
      (a)
      \includegraphics[scale=0.7]{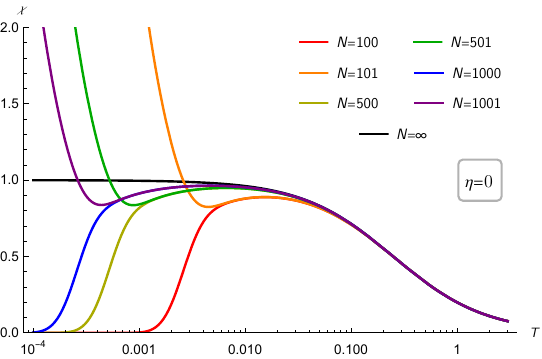}
    \end{minipage}
    \hspace{15mm}
    \begin{minipage}{0.4\textwidth}
      (b)
      \includegraphics[scale=0.7]{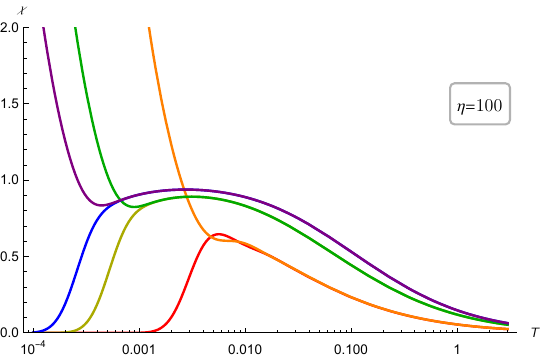}
    \end{minipage}
    \caption{Magnetic susceptibility as a function of temperature for values of
    the number of particles $N=100$ (red), $101$ (orange), $N=500$ (yellow), $501$
    (green), $N=1000$ (blue), $1001$ (purple) for the antiferromagnetic case.
    The values of the parameter $q=e^{\eta/N}$ are: (a) $\eta=0$ and (b) $\eta=100$.}
    \label{qkskj1/2transitionchi}
  \end{figure}

\begin{figure}[H]
    \begin{minipage}{0.4\textwidth}
      (a)
      \includegraphics[scale=0.7]{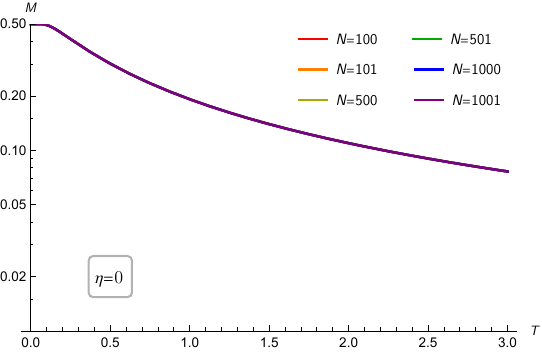}
    \end{minipage}
    \hspace{15mm}
    \begin{minipage}{0.4\textwidth}
      (b)
      \includegraphics[scale=0.7]{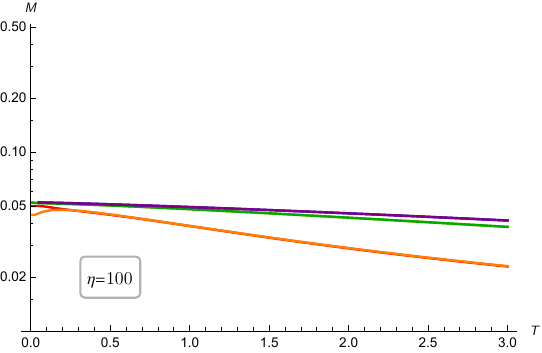}
    \end{minipage}
    \caption{Magnetisation as a function of temperature for values of the number
    of particles $N=100$ (red), $101$ (orange), $N=500$ (yellow), $501$ (green),
    $N=1000$ (blue), $1001$ (purple), for the ${h\gamma=1}$ antiferromagnetic
    case. The values of the parameter $q=e^{\eta/N}$ are: (a) $\eta=0$ and (b) $\eta=100$.}
    \label{qkskj1/2transitionm}
  \end{figure}

As in the undeformed scenario, we can study the thermodynamic limit in the deformed case {by numerical evaluation of the exact partition function~\eqref{partition}}. In addition to the modifications observed for small $N$, when the value of the deformation parameter increases, the convergence between contiguous $N$ (with different parity) requires a higher temperature. {Similar to the analysis of magnetisation jumps, the figures representing the thermodynamic limit (Figs.~\ref{qkskj1/2transitioncv}, \ref{qkskj1/2transitionchi}, and \ref{qkskj1/2transitionm}) employ moderate system sizes ($N$ up to 1001). Within the considered temperature range, these sizes are already sufficient to achieve full convergence, ensuring that the system has effectively reached the thermodynamic limit while avoiding excessive numerical noise.}

First, as in the undeformed case, the maxima are shifted to the left for increasing values of $N$ (for the same $q$). In particular, in Fig.~\ref{qkskj1/2transitioncv} (b), several bumps can be observed for both the even and odd $N$ cases, which become less sharp as the temperature increases. These are due to transitions between consecutive $J$ energy levels. These transitions overlap in the undeformed case; however, as explained previously, the energy required to reach the excited levels increases when deformation is introduced. By plotting the most probable levels, it can be seen that these bumps are due to the transitions between consecutive $J$-energetic levels. Every transition occurs at higher temperatures as $\eta$ increases because the difference between each pair of energy levels is also greater. These bumps are more pronounced and smaller with increasing values of the deformation parameter because the deformation becomes more visible at high temperatures. This occurs because the deformation increases the energy differences between levels, as seen in \cite{Ballesteros:2025cia}, and thus the transitions are visible at higher temperatures. The decrease in the magnitude can be understood from the fact that although the first {ground} states have less energy when the deformation parameter is increased, the energy of the next excited levels becomes exponentially larger. Therefore, the lowest-energy states become more likely, and the total energy of the system is closer to the energy of the lowest level. This implies that the specific heat exhibits smaller bumps for larger values of $\eta$.

In the deformed case, the most probable levels for any number of particles are the same as in the undeformed scenario ($J=0,1$ for $N$ even and $J=1/2, 3/2$ for $N$ odd). Consequently, we can also get the analytic expressions of specific heat in the deformed case
  \begin{align}
    C_{V,q}^{e,TL} & {\approx}\frac{9 (q+1)^{2}e^{\frac{q^{3/2}+\frac{1}{\sqrt{q}}}{2 N T(1-q)}}}{4 N^{3}q T^{2}\left(9e^{\frac{q^{3/2}}{2 N T(1-q)}}+ e^{\frac{1}{2NT\sqrt{q}(1-q)}}\right)^{2}}= \notag                                                                                                                     \\
                   & =C_{V}^{e,TL}+\frac{9 e^{\frac{1}{N T}}\left(18 N T+e^{\frac{1}{N T}}(2 N T-1)+9\right)}{8 N^{4}T^{3}\left(e^{\frac{1}{N T}}+9\right)^{3}}(q-1)^{2}+O\left((q-1)^{3}\right),                                                                                                                     \\
    C_{V,q}^{o,TL} & {\approx}\frac{\left(q^{2}+q+1\right)^{2}e^{\frac{(q+1) \left(q+\sqrt{q}+1\right)^2}{2 N \left(\sqrt{q}+1\right)^2 q T}}}{N^{3}q^{2}T^{2}\left( e^{\frac{\left(1-q^{3/2}\right) \left(1-q^{5/2}\right)}{2 N (q-1)^2 q T}}+4 e^{\frac{q+\sqrt{q}+1}{2 N \left(\sqrt{q}+1\right)^2 T}}\right)^{2}}= \notag \\
                   & =C_{V}^{o,TL}+\frac{3 e^{\frac{3}{N T}}(4 N T-3)+12 e^{\frac{3}{2 N T}}(4 N T+3)}{2N^{4}T^{3}\left(e^{\frac{3}{2 N T}}+4\right)^{3}}(q-1)^{2}+O\left((q-1)^{3}\right).
  \end{align}

For the magnetic susceptibility, the convergence observed in the undeformed case is attenuated, as shown in Fig. \ref{qkskj1/2transitionchi}. As deformation is introduced, the curves require a higher temperature to converge and practically do so for a zero value of the susceptibility. Moreover, as in the undeformed case, the curves shift to the left for increasing values of $N$ (for the same $q$).

As in the undeformed case, the most probable levels of magnetic susceptibility are the same as for the specific heat ($J=0,1$ for $N$ even and $J=1/2, 3/2$ for $N$ odd). This fact allows us to obtain analytic expressions of the magnetic susceptibility in the deformed case
  \begin{align}
    \chi^{e,TL}_{q} & {\approx}\frac{6}{N T (e^{\frac{q+1}{2 N \sqrt{q} T}}+9)}=\chi^{e,TL}-\frac{3 e^{\frac{1}{N T}}}{4 N^{2}T^{2}\left(e^{\frac{1}{N T}}+9\right)^{2}}(q-1)^{2}+O\left((q-1)^{3}\right),                                                                                                                    \\
    \chi^{o,TL}_{q} & {\approx}\frac{e^{\frac{q^4+1}{2 N (q-1)^2 q T}}+20e^{\frac{q^2+1}{2 N (q-1)^2 T}}}{4 N T \left(e^{\frac{q^4+1}{2 N (q-1)^2 q T}}+4e^{\frac{q^2+1}{2 N (q-1)^2 T}}\right)}=\chi^{o,TL}-\frac{2 e^{\frac{3}{2 N T}}}{N^{2}T^{2}\left(e^{\frac{3}{2 N T}}+4\right)^{2}}(q-1)^{2}+O\left((q-1)^{3}\right).
  \end{align}

For magnetisation, the most probable levels are not the same as those obtained for the specific heat and magnetic susceptibility because an external magnetic field is applied. In fact, they are not even the same levels for every value of the deformation parameter. To describe them, we obtained an equation that allows us to obtain the most probable levels from the deformation parameter and the number of particles. This is obtained by minimizing the energy expression with an external magnetic field \eqref{eq:qenergy} (for {$h\gamma=1$} and $m=J$) as a function of $J$
  \begin{equation}
    J_{MP}=m_{MP}=\frac{\mathop{N\cdot \mathrm{arcsinh}}\left(\frac{2 N^{2}(\cosh
    (\eta/N)-1)}{\eta }\right)}{\eta }-\frac{1}{2}. \label{eq:MostProbable}
  \end{equation}

The two most probable energy states are those with angular momentum being the closest integers for even $N$ and the closest semi-integer for odd $N$. From them, we would also be able to write the analytical approximation carried out for the undeformed case. However, because the most probable states depend on $N$, we cannot write a generic expression as we did in the specific heat and the magnetic susceptibility.
  
Owing to the dependence of the most probable levels on the number of particles and the value of the deformation parameter, the deformed magnetisation at $T=0$ has different values for different $q$ and $N$, as shown in Fig. \ref{qkskj1/2transitionm} (b). These magnetisation values at zero temperature can be calculated using \eqref{eq:MostProbable}: $M(T=0)=m_{MP}/N$. We show the computation of the most probable states (magnetisation at zero temperature) for three deformation values in Table~\ref{table:nmjmp}.
  \begin{table}[H]
    \centering
    \begin{tabular}{c c c c}
      \hline
      $N$  & |$J_{MP},m_{MP}\rangle_{q=e^{10/N}}$ & |$J_{MP},m_{MP}\rangle_{q=e^{50/N}}$ & |$J_{MP},m_{MP}\rangle_{q=e^{100/N}}$ \\
      \hline
      100  & |$29,29\rangle$                      & |$9,9\rangle$                        & |$5,5\rangle$                         \\
      \hline
      101  & |$29.5,29.5\rangle$                  & |$8.5,8.5\rangle$                    & |$4.5,4.5\rangle$                     \\
      \hline
      500  & |$149,149\rangle$                    & |$46,46\rangle$                      & |$26,26\rangle$                       \\
      \hline
      501  & |$149.5,149.5\rangle$                & |$45.5,45.5\rangle$                  & |$26.5,26.5\rangle$                   \\
      \hline
      1000 & |$299,299\rangle$                    & |$92,92\rangle$                      & |$52,52\rangle$                       \\
      \hline
      1001 & |$299.5,299.5\rangle$                & |$91.5,91.5\rangle$                  & |$52.5,52.5\rangle$                   \\
      \hline
    \end{tabular}
    \caption{Data for different number of particles of the most probable states
    in presence of an external magnetic field {$h\gamma=1$}.}
    \label{table:nmjmp}
  \end{table}

  The curves associated with an odd number of particles demonstrate an initial increase
  at lower temperatures, followed by a subsequent decrease. This behavior can be
  explained by the energy distribution, where, for an odd number of particles,
  the first excited state corresponds to the next consecutive magnetic quantum
  number $m$. This leads to a slight increase in magnetisation. In contrast, for
  an even number of particles, the first excited state corresponds to the
  previous consecutive magnetic quantum number $m$, resulting in a decrease in the
  magnetisation.

  In addition, the decrease in magnetisation becomes more gradual as the
  deformation parameter increases. This is again due to the fact that the deformed
  systems require more energy to reach the excited states. Consequently, the magnetisation
  behaviour is dominated by the most probable states.

\begin{figure}[H]
    \begin{minipage}{0.4\textwidth}
      {(B)}
      \includegraphics[scale=0.7]{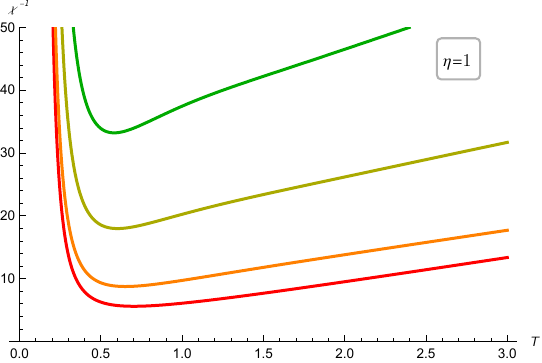}
    \end{minipage}
    \hspace{15mm}
    \begin{minipage}{0.4\textwidth}
      \includegraphics[scale=0.7]{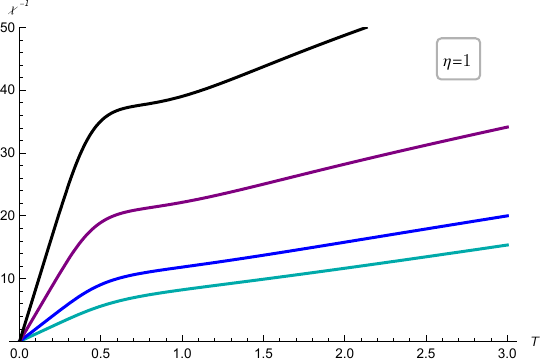}
    \end{minipage}
    \caption{{Inverse of magnetic susceptibility as a function of temperature for
    values of (A) the even number of particles $N=2$ (red), $4$ (orange), $10$ (yellow)
    and $20$ (green) and (B) the odd number of particles $N=3$ (cyan), $5$ (blue),
    $11$ (purple), and $21$ (black), and of the parameter (A) $\eta=0$ and (B) $\eta=1$, for the antiferromagnetic case.}}
    \label{fig:curieantid}
  \end{figure}

  We can now obtain the Curie temperature. We begin by representing the inverse
  of the magnetic susceptibility in Fig.~\ref{fig:curieantid} for different values
  of $q$.
  \begin{figure}[H]
    \begin{minipage}{0.4\textwidth}
      (A)
      \includegraphics[scale=0.7]{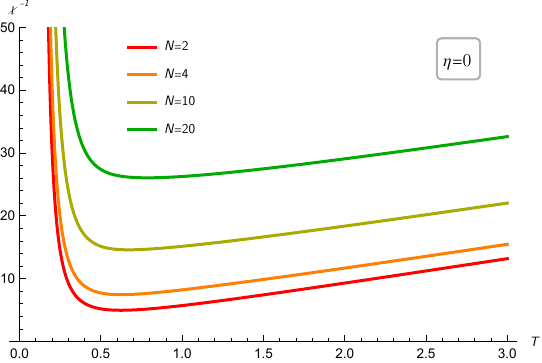}
    \end{minipage}
    \hspace{15mm}
    \begin{minipage}{0.4\textwidth}
      \includegraphics[scale=0.7]{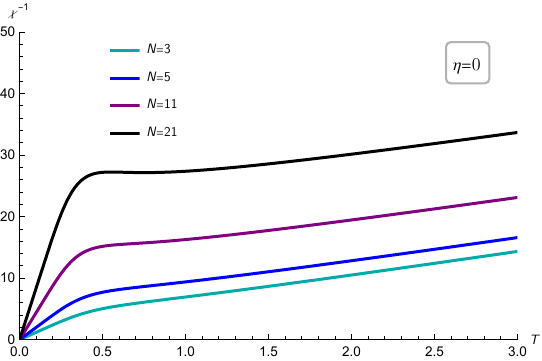}
    \end{minipage}
  \end{figure}

  Taking the analytic expressions of the most probable states we find
  \begin{equation}
    \frac{1}{\chi^{e}}{\approx}\frac{T \left((N+4) e^{\frac{q+1}{2 \sqrt{q} T}}+9 N\right)}{6}
    ,
  \end{equation}
  \begin{equation}
    \frac{1}{\chi^{o}}{\approx}\frac{4 N T \left((N+5) e^{\left(\frac{\left(1-q^{3/2}\right)
    \left(1-q^{5/2}\right)}{2 (q-1)^2 q T}\right)}+4 (N-1) e^{\frac{q+\sqrt{q}+1}{2
    \left(\sqrt{q}+1\right)^2 T}}\right)}{(N+5) e^{\left(\frac{\left(1-q^{3/2}\right)
    \left(1-q^{5/2}\right)}{2 (q-1)^2 q T}\right)}+20 (N-1) e^{\frac{q+\sqrt{q}+1}{2
    \left(\sqrt{q}+1\right)^2 T}}},
  \end{equation}
  The even susceptibility does not diverge, so it does not make sense to
  calculate the Curie temperature in this case. In the odd case, the Curie
  temperature is $T_{C}=0$ and the Curie constant is $C=\frac{1}{4N}$, exactly as
  in the undeformed case.

{
\subsection{Finite-size scaling}

In the antiferromagnetic regime, the finite-size behavior of the $q$-deformed Kittel--Shore model is strongly influenced by the parity of the system size at low temperatures. Even and odd chains have different ground-state structures, and this difference is reflected in the location and height of the specific-heat maximum for finite $N$. Nevertheless, when the thermodynamic observables are analyzed at the temperature $T_{\max}$ at which the specific heat is maximal, both parity sectors display the same qualitative asymptotic scenario: the specific-heat peak decays as $N^{-1}$, while the magnetic susceptibility approaches a finite, size-independent limit. In this sense, although parity determines different microscopic low-energy sectors, it does not modify the overall structure of the finite-size scaling laws in the thermodynamic limit.

This behavior is robust over a broad range of deformation parameters $\eta$, provided that the numerically stable condition $\eta/N \ll 1$ is satisfied. Within this regime, the data show that the deformation does not affect the asymptotic scaling exponents, and only parity distinguishes the limiting amplitudes. The results are independent of $\eta$, being equally valid for the undeformed case and for $\eta=100$, which stands in clear contrast to the ferromagnetic case.

\subsubsection{Even number of spins}
In the thermodynamic limit, the scaling variable is $s=NT$. The specific-heat maximum and its height scale as
\[
T_{\max}(N)\sim \frac{0.296}{N}, \qquad C_V^{\max}(N)\sim \frac{2.115}{N},
\]
yielding a finite limit for $NC_V^{\max}$. The magnetic susceptibility at $T_{\max}$ remains finite, $\chi(T_{\max})\to 0.53365\ldots$.

Table~\ref{tab:even-eta} confirms this across different $\eta$. For $\eta/N\ll 1$, $NT_{\max}$ and $NC_V^{\max}$ stabilize rapidly. A log-log analysis yields scaling exponents $p_{C_V}=-1$ and $p_{\chi}=0$.

\begin{table}[H]
\centering
\begin{tabular}{c c c c c}
\hline
$\eta$ & $N$ & $NT_{\max}$ & $NC_V^{\max}$ & $\chi(T_{\max})$ \\
\hline
\multirow{8}{*}{$0$} 
        & $10^3$          & $0.2966$ & $2.1105$  & $0.533757$ \\
        & $3\cdot 10^3$   & $0.2965$ & $2.11383$ & $0.534573$ \\
        & $6\cdot 10^3$   & $0.2965$ & $2.11466$ & $0.534852$ \\
        & $10^4$          & $0.2965$ & $2.115$   & $0.534963$ \\
        & $10^5$          & $0.296$  & $2.11544$ & $0.533635$ \\
        & $3\cdot 10^5$   & $0.296$  & $2.11547$ & $0.533646$ \\
        & $6\cdot 10^5$   & $0.296$  & $2.11548$ & $0.533649$ \\
        & $10^6$          & $0.296$  & $2.11548$ & $0.53365$  \\
\hline
\multirow{8}{*}{$50$} 
        & $10^3$          & $0.2967$ & $2.11007$ & $0.533572$ \\
        & $3\cdot 10^3$   & $0.2965$ & $2.11378$ & $0.53452$  \\
        & $6\cdot 10^3$   & $0.2965$ & $2.11465$ & $0.534838$ \\
        & $10^4$          & $0.2965$ & $2.11499$ & $0.534958$ \\
        & $10^5$          & $0.296$  & $2.11544$ & $0.533635$ \\
        & $3\cdot 10^5$   & $0.296$  & $2.11547$ & $0.533646$ \\
        & $6\cdot 10^5$   & $0.296$  & $2.11548$ & $0.533649$ \\
        & $10^6$          & $0.296$  & $2.11548$ & $0.53365$  \\
\hline
\multirow{8}{*}{$100$} 
        & $10^3$          & $0.2967$ & $2.10881$ & $0.532138$ \\
        & $3\cdot 10^3$   & $0.2965$ & $2.11364$ & $0.53436$  \\
        & $6\cdot 10^3$   & $0.2965$ & $2.11461$ & $0.534798$ \\
        & $10^4$          & $0.2965$ & $2.11498$ & $0.534944$ \\
        & $10^5$          & $0.296$  & $2.11544$ & $0.533635$ \\
        & $3\cdot 10^5$   & $0.296$  & $2.11547$ & $0.533646$ \\
        & $6\cdot 10^5$   & $0.296$  & $2.11548$ & $0.533649$ \\
        & $10^6$          & $0.296$  & $2.11548$ & $0.53365$  \\
\hline
\end{tabular}
{\caption{Asymptotic results for even $N$ and several deformation parameters $\eta$, restricted to the numerically stable regime $\eta/N \ll 1$.}
\label{tab:even-eta}}
\end{table}

\subsubsection{Odd number of spins}

For odd $N$, the specific-heat maximum shifts to $s_{\star}^{(\mathrm{odd})}=1$, giving
\[
T_{\max}(N)\sim \frac{1}{N}, \qquad C_V^{\max}(N)\sim \frac{3}{2N}.
\]
The magnetic susceptibility at $T_{\max}$ converges to $\chi(T_{\max})\to 0.91666\ldots \approx 11/12$.

Table~\ref{tab:odd-eta} shows the odd sector shares the same scaling laws ($p_{C_V}=-1$, $p_{\chi}=0$), differing only in asymptotic constants. Data for different $\eta$ collapse toward this behavior as $N$ grows.

\begin{table}[H]
\centering
\begin{tabular}{c c c c c}
\hline
$\eta$ & $N$ & $NT_{\max}$ & $NC_V^{\max}$ & $\chi(T_{\max})$ \\
\hline
\multirow{8}{*}{$0$} 
        & $10^3+1$        & $0.85$   & $1.48913$ & $0.898619$ \\
        & $3\cdot 10^3+1$ & $0.9182$ & $1.49609$ & $0.908032$ \\
        & $6\cdot 10^3+1$ & $0.9612$ & $1.49796$ & $0.912668$ \\
        & $10^4+1$        & $0.9926$ & $1.49874$ & $0.915652$ \\
        & $10^5+1$        & $1.0$    & $1.49982$ & $0.916629$ \\
        & $3\cdot 10^5+1$ & $1.0$    & $1.4999$  & $0.916656$ \\
        & $6\cdot 10^5+1$ & $1.0$    & $1.49992$ & $0.916663$ \\
        & $10^6+1$        & $1.0$    & $1.49992$ & $0.916665$ \\
\hline
\multirow{8}{*}{$50$} 
        & $10^3+1$        & $0.834$  & $1.48659$ & $0.895306$ \\
        & $3\cdot 10^3+1$ & $0.9128$ & $1.49578$ & $0.907316$ \\
        & $6\cdot 10^3+1$ & $0.9584$ & $1.49788$ & $0.912367$ \\
        & $10^4+1$        & $0.9911$ & $1.49871$ & $0.915498$ \\
        & $10^5+1$        & $1.0$    & $1.49982$ & $0.916629$ \\
        & $3\cdot 10^5+1$ & $1.0$    & $1.4999$  & $0.916656$ \\
        & $6\cdot 10^5+1$ & $1.0$    & $1.49992$ & $0.916663$ \\
        & $10^6+1$        & $1.0$    & $1.49992$ & $0.916665$ \\
\hline
\multirow{8}{*}{$100$} 
        & $10^3+1$        & $0.8006$ & $1.47937$ & $0.887109$ \\
        & $3\cdot 10^3+1$ & $0.8988$ & $1.49485$ & $0.905368$ \\
        & $6\cdot 10^3+1$ & $0.951$  & $1.49763$ & $0.911549$ \\
        & $10^4+1$        & $0.9864$ & $1.49861$ & $0.915053$ \\
        & $10^5+1$        & $1.0$    & $1.49981$ & $0.916628$ \\
        & $3\cdot 10^5+1$ & $1.0$    & $1.4999$  & $0.916656$ \\
        & $6\cdot 10^5+1$ & $1.0$    & $1.49992$ & $0.916663$ \\
        & $10^6+1$        & $1.0$    & $1.49992$ & $0.916665$ \\
\hline
\end{tabular}
{\caption{Asymptotic results for odd $N$ and several deformation parameters $\eta$, restricted to the numerically stable regime $\eta/N \ll 1$.}
\label{tab:odd-eta}}
\end{table}

Taken together, these results show that the antiferromagnetic regime exhibits a parity-dependent low-energy structure but a remarkably simple thermodynamic scaling pattern. Even and odd sequences are governed by different universal amplitudes for $T_{\max}$, $C_V^{\max}$, and $\chi(T_{\max})$, yet both sectors share the same asymptotic exponents: the specific-heat maximum decays as $N^{-1}$, while the susceptibility saturates to a finite limit.}

{
\section{Deformation for \texorpdfstring{$|q|=1$}{|q|=1} and comparison with an experimental model}
\label{sec:complex}

In this section, we consider a generic unit-circle deformation parameter,
\begin{equation}
    q = e^{i\theta}, \qquad \frac{\theta}{\pi} \notin \mathbb{Q},
    \label{eq:q_phase}
\end{equation}
restricting our analysis to values of $q$ that are not roots of unity.

The purpose of this section is twofold. First, we show the effects of a complex
$q$-deformation. Since all the previous analytical results are completely valid and since we are interested in $N=4$ in the following, we restrict ourselves to this particular case.    Second, we compare the
{thermal} predictions of the $q$-deformed model with the microscopic
Hamiltonian used to describe the intratetrahedral interaction of the family of quantum magnets
Cu$_2$Te$_2$O$_5$X$_2$ ($X=$ Cl, Br) \cite{jaglicic2006magnetic}. These
materials are formed by distorted tetrahedral clusters of
$S=1/2$ Cu$^{2+}$ spins, with dominant antiferromagnetic interactions inside
the tetrahedra and additional intertetrahedral couplings. The comparison
shows that the phase $\theta$ of the deformation parameter can be used as an effective parameter for describing the intratetrahedral interaction even for a small temperature.

\subsection{Complex \texorpdfstring{$q$}{q}-deformation as an effective algebraic renormalization}
\label{subsec:q_complex}

For $|q|=1$, the deformation introduces {an oscillatory behavior of the energy splitting between spin multiplets}. When $q$ is on the unit circle, $q=e^{i\theta}$ with $\theta \in [0, 2\pi)$, the corresponding $q$-number can be written as 
\begin{equation}
    [n]_q
    =
    \frac{q^{n/2}-q^{-n/2}}{q^{1/2}-q^{-1/2}}
    =
    \frac{\sin(n\theta/2)}{\sin\theta/2}.
    \label{eq:q_number_complex}
\end{equation}

In order to {illustrate} the oscillatory behavior of this kind of deformation and its differences with respect to the real case, we show, for $N=4$ and for different values of the deformation parameter, the plots of eigenvalues, specific heat, magnetic susceptibility, and magnetisation. Note that the eigenvalues and {thermal} observables vary continuously as a function of $q$,
even when $q$ approaches {low-order} root-of-unity values on the unit circle. Nevertheless,
as mentioned previously, at these special points the representation theory of $U_q(\mathfrak{su}(2))$
completely changes, so these values,
while present in the plots, are excluded from the physical interpretation.

\begin{figure}[H]
    \centering
    \includegraphics[scale=0.9]{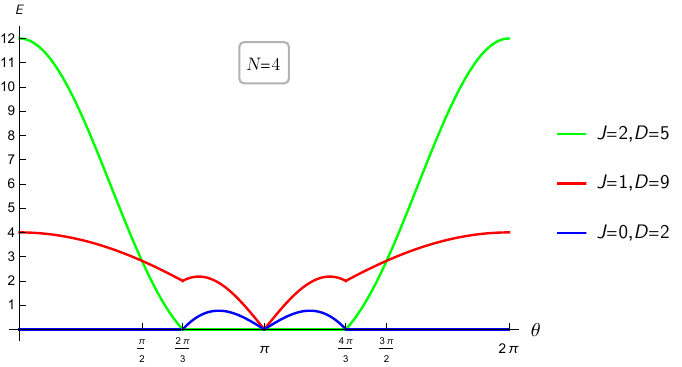}
    {\caption{Energy levels (in the units $|I|/4$) and their degeneracies for $N=4$ spins with $j=1/2$ coupled antiferromagnetically without an external magnetic field $(h=0)$ as a function of the deformation parameter $\theta$ (where $q=e^{i\theta}$) displaced from the {ground state}. 
    {Note that $(1/2)^{\otimes 4} = 2 \oplus 1^{\oplus 3} \oplus 0^{{\oplus}2}$.}}
    \label{qkskj1/2eigenAF}}
  \end{figure}
The plot illustrates the evolution of the energy levels for different total spin sectors $J$, with their respective degeneracies $D$. The spectrum displays a clear reflection symmetry about $\theta = \pi$, where a total collapse of the energy gaps occurs. {A reflection of the phase angle about $\theta = \pi$ (i.e., $\theta \to 2\pi - \theta$) is algebraically equivalent to parameter inversion $q \to q^{-1}$ since $e^{i(2\pi - \theta)} = e^{2\pi i} e^{-i\theta} = 1 \cdot e^{-i\theta} = q^{-1}$, which directly establishes the reflection symmetry of the energy spectrum by virtue of the previously proven invariance in Sec. \ref{invariance}:
\begin{equation}
    E_{N,J,m}^{q(2\pi - \theta)} = E_{N,J,m}^{e^{i(2\pi - \theta)}} = E_{N,J,m}^{e^{2\pi i} e^{-i\theta}} = E_{N,J,m}^{q^{-1}(\theta)} = E_{N,J,m}^{q(\theta)}.
\end{equation}}
A key feature of this system is the spectral inversion observed in the region $\theta \in [2\pi/3, 4\pi/3]$; here, the $J=2$ sector, which constitutes the most excited state in the undeformed limit ($\theta = 0, 2\pi$), becomes the ground state. {As shown in Appendix~\ref{app:crossings}, exact energy crossings between magnetic sectors occur when $q$ is a root of unity. While the eigenvalues remain continuous in $q$---being roots of a Laurent-polynomial characteristic equation in $q^{1/2}$---the underlying representation theory of $U_q(\mathfrak{su}(2))$ becomes non-semisimple. Crucially, this does not invalidate the spectral framework, but instead produces enhanced state degeneracies. A full investigation of these root-of-unity algebraic structures is left for future work.}


\subsection{{Thermal} properties}
We have numerically computed the {thermal} properties for several values of the deformation parameter $\theta \in \{0.5, 1, 2, 3\}$. These choices represent distinct spectral regimes, as they fall within different intervals where the distribution of eigenvalues changes: $0<0.5<1<\pi/2$, $\pi/2<2<2\pi/3$ and $2\pi/3<3<2\pi$. {We represent the specific heat for these deformation values as a function of temperature in Fig. \ref{cvcom}.}
\begin{figure}[H]
    \centering
    \includegraphics[scale=0.9]{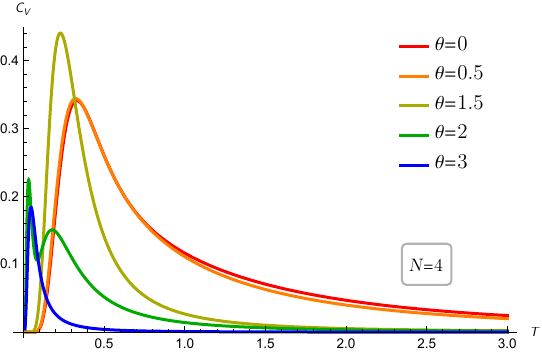}
    \caption{Specific heat as a function of temperature for value of the number of particles $N=4$ and of the parameter $\theta=0$ (red), $\theta=0.5$ (orange), $\theta=1.5$ (darker yellow), $\theta=2$ (green) and $\theta=3$ (blue) (where $q=e^{i\theta}$) for the antiferromagnetic case.}
    \label{cvcom}
\end{figure}

The deformation parameter $\theta$ acts as a continuous tuning mechanism for the energy spectrum. As $\theta \to \pi/2$, the energy gap separating the ground state from the first excited states gradually shrinks. From a {thermal} perspective, this reduction in the energy difference implies that significantly less thermal energy is required for the system to undergo transitions between these states. This physical mechanism explains why the specific heat peak, which precisely marks the characteristic temperature scale of such transitions, shifts toward progressively lower temperatures as the eigenvalues approach one another. 

Furthermore, the overall peak magnitude is strongly dependent on the density of states. For instance, the peak observed for $\theta=1.5$ is noticeably higher than the one for $\theta=0.5$. This occurs because the level compression near $\pi/2$ effectively concentrates the thermal absorption into a narrower energy band. In contrast, for $\theta=2$, the spectral restructuring that takes place past $\pi/2$ gives rise to two distinct energy scales, ultimately producing a characteristic double peak. Finally, for $\theta=3$, as the system approaches $\pi$, the eigenvalues crowd together to such an extent that sharp transitions begin to vanish. The thermal excitation diffuses across a set of nearly degenerate levels, smoothing out the specific heat curve and substantially lowering its maximum value.
\begin{figure}[H]
    \centering
    \includegraphics[scale=0.9]{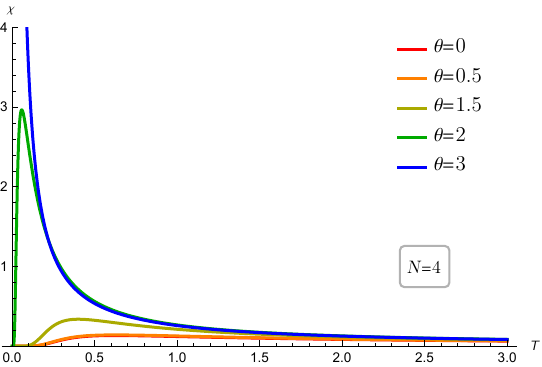}
    \caption{Magnetic susceptibility as a function of temperature for value of the number of particles $N=4$ and of the parameter $\theta=0$ (red), $\theta=0.5$ (orange), $\theta=1.5$ (darker yellow), $\theta=2$ (green) and $\theta=3$ (blue) (where $q=e^{i\theta}$) for the antiferromagnetic case. The peaks are shifted toward higher temperatures and the maximum susceptibility is reached for smaller values of $\theta$.}
    \label{chicom}
\end{figure}

{The magnetic susceptibility at the same specific deformation values is plotted
against temperature in Fig. \ref{chicom}. The magnetic susceptibility} directly reflects the consequences of spectral compression. Near $\pi/2$, the significantly reduced energy gaps enhance the overall magnetic sensitivity of the system, which translates into increased peaks at moderate temperatures. In the case of $\theta=2$, a distinct and sharp low-temperature peak appears, driven by the fact that the $J=2$ state is located very close to the ground state. However, for $\theta=3$, the $J=2$ state actually becomes the new ground state. Under these conditions, the susceptibility simply decays monotonically, as the available thermal excitation is no longer sufficient to reach higher-moment states.

\begin{figure}[H]
    \centering
    \includegraphics[scale=0.9]{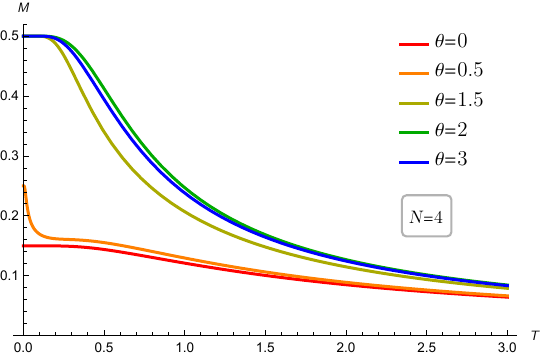}
    \caption{Magnetisation as a function of temperature for value of the number of particles $N=4$ and of the parameter $\theta=0$ (red), $\theta=0.5$ (orange), $\theta=1.5$ (darker yellow), $\theta=2$ (green) and $\theta=3$ (blue) (where $q=e^{i\theta}$) for the $h\gamma=1$ antiferromagnetic case.}
    \label{magcom}
\end{figure}

{Finally, we represent the magnetisation for the aforementioned deformation values as a function of temperature in Fig. \ref{magcom}.} For the chosen external magnetic field, the ground state is degenerate and shared by both the $J=m=0$ and $J=m=1$ states in the undeformed limit, $\theta=0$ (as shown in Fig. \ref{qkskj1/2mp} (a)). However, as the deformation parameter $\theta$ increases, this degeneracy is broken. The ground level becomes solely occupied by the $J=m=1$ state within the interval $0 < \theta < \theta_c$. This upper limit corresponds to the critical deformation where a level crossing occurs, determined analytically by the root of $2\cos(\theta)\cos(\theta/2)-1=0$, yielding the value 
\begin{equation}
    \theta_c = 2\arccos\left( \sqrt[3]{\frac{\sqrt{57}}{72} + \frac{1}{8}} + \frac{1}{6\sqrt[3]{\frac{\sqrt{57}}{72} + \frac{1}{8}}} \right) \approx 0.9701.
\end{equation}

For deformations beyond this critical point ($\theta_c < \theta < 2\pi$), the ground state is exclusively occupied by the $J=m=2$ state. Consequently, at a fixed deformation such as $\theta=0.5$ (which lies in the first regime), the system exhibits a non-zero magnetization of $0.25$ at $T=0$, reflecting the exclusive population of the $J=m=1$ ground state. As the temperature increases, the magnetization rapidly decays because the first accessible thermal excitation is the non-magnetic $J=m=0$ singlet.

For the higher deformation values ($\theta=1.5, 2$, and $3$), the ground state is exclusively given by $J=m=2$, which means the system is strongly magnetized even at $T=0$. As thermal effects are introduced, the magnetisation shows a staggered, step-like decay. This occurs because the system loses its magnetisation hierarchically across a set of fragmented energy levels with very close spin projections. Ultimately, the phase $\theta$ acts to completely redefine the stable configuration of the system at low temperatures.

\subsection{Microscopic tetrahedral Hamiltonian}
\label{subsec:tetra_hamiltonian}

{To clarify the geometric and topological interpretation of the $N=4$ spin cluster, it is instructive to examine the structure of its pairwise interactions. In the undeformed limit ($q=1$), the system exhibits isotropic all-to-all coupling among all $\binom{4}{2}=6$ spin pairs, each with identical strength $I$, reflecting the full spatial symmetry of a regular tetrahedron.}

{Introducing the $q$-deformation ($q \neq 1$) breaks this permutation symmetry, as the underlying $U_q(\mathfrak{su}(2))$ quantum algebra splits the six interaction channels into two inequivalent classes. Four \emph{primary} couplings---$(1,2)$, $(2,3)$, $(3,4)$, and $(4,1)$---correspond to the four side edges of a distorted tetrahedron, or equivalently, the four nearest-neighbor bonds of a closed ring. The remaining two \emph{secondary} couplings---$(1,3)$ and $(2,4)$---correspond to the two opposite edges of the tetrahedron, or equivalently, the two next-nearest-neighbor bonds across the ring's diameter.}

{This splitting admits two dual, structurally equivalent geometric pictures. Geometrically, it can be viewed as a spatially distorted tetrahedron in which four edges are equally deformed relative to the remaining two. Alternatively, in a 1D/2D periodic setting, it describes a ring (or square) array of $N=4$ spins with two effective interaction ranges: nearest-neighbor couplings along the perimeter and next-nearest-neighbor couplings across the diameter. Importantly, $q$ does not represent a physical displacement of the spin sites; rather, it modulates the relative anisotropy between these two interaction classes within the algebraic framework.}

{This theoretical picture maps naturally onto physical compounds characterized by structurally distorted spin tetrahedra.} The geometrical frustration in Cu$_2$Te$_2$O$_5$X$_2$ originates from the
antiferromagnetic couplings among four Cu$^{2+}$ spins arranged in a distorted
tetrahedron. The intratetrahedral Hamiltonian of an isolated cluster
($N=4$) can be written as \cite{jaglicic2006magnetic}
\begin{equation}
    \mathcal{H}_{\mathrm{tetra}}
    =
    J_1\left(
        \mathbf{S}_{1}\cdot\mathbf{S}_{3}
        +
        \mathbf{S}_{2}\cdot\mathbf{S}_{4}
    \right)
    +
    J_2\left(
        \mathbf{S}_{1}\cdot\mathbf{S}_{2}
        +
        \mathbf{S}_{2}\cdot\mathbf{S}_{3}
        +
        \mathbf{S}_{3}\cdot\mathbf{S}_{4}
        +
        \mathbf{S}_{4}\cdot\mathbf{S}_{1}
    \right),
    \label{eq:H_tetra}
\end{equation}
where $J_1$ corresponds to the two longer diagonal bonds and $J_2$ to the four
shorter bonds forming a ring-like structure. In the idealized limit
$J_1=J_2$, all bonds would be equivalent; however, in the real compounds the
tetrahedron is distorted, and the ratio $J_2/J_1$ becomes a direct measure of
the degree of intracluster frustration.

The complete microscopic description considered in Ref.~\cite{jaglicic2006magnetic}
also includes intertetrahedral couplings along the chain direction and a
mean-field coupling between neighboring chains. In that approach the total
Hamiltonian is schematically decomposed as
\begin{equation}
    \mathcal{H}
    =
    \mathcal{H}_{\mathrm{chain}}
    +
    \mathcal{H}_{\mathrm{MFA}},
    \label{eq:H_total_micro}
\end{equation}
where $\mathcal{H}_{\mathrm{chain}}$ contains both the intratetrahedral
couplings $J_1,J_2$ and the intertetrahedral couplings $J_3,J_4$, while
$\mathcal{H}_{\mathrm{MFA}}$ accounts for the coupling $J_c$ between
neighboring chains within a mean-field approximation. In the present work we
focus on the isolated tetrahedral cluster as the fundamental microscopic unit,
and use the $q$-deformation to reproduce its effective {thermal} response.

In the effective {thermal} model, the microscopic frustration encoded by
the pair $(J_1,J_2)$ is replaced by the same set of algebraic parameters:
an effective exchange scale $I$ and a deformation phase $\theta$. 
The central idea is that the complex phase $\theta$ acts as a compact algebraic
parameter that reproduces the cumulative {thermal} effect of the distorted
tetrahedral geometry. In this sense, the deformation performs an effective
coarse-graining:
\begin{equation}
    (J_1,J_2)
    \quad
    \longrightarrow
    \quad
    \bigl(I,\theta\bigr).
    \label{eq:coarse_graining}
\end{equation}
This mapping should not be interpreted as an operator identity between the
microscopic Hamiltonian and the $q$-deformed Hamiltonian. Instead, it is a
{thermal} equivalence: both descriptions generate nearly identical
{global} observables over the temperature and field ranges considered.

\subsubsection{Numerical comparison with the isolated tetrahedron}
\label{subsubsec:numerical_comparison}

To test the validity of the effective description, we fitted the
$q$-deformed partition function to exact-diagonalization data for an isolated
tetrahedron ($N=4$), using experimentally motivated parameters for the
compounds
\[
    \mathrm{Cu_2Te_2O_5Cl_2},\qquad
    \mathrm{Cu_2Te_2O_5ClBr},\qquad
    \mathrm{Cu_2Te_2O_5Cl_{1.5}Br_{0.5}},\qquad
    \mathrm{Cu_2Te_2O_5Br_2}.
\]
The fit was performed simultaneously for the specific heat $C_{V}/T$, the magnetic
susceptibility $\chi$, and the magnetisation $M$ in an external magnetic
field. This is done for a range of temperature between 100 and 300 K, for which the $q$-deformation better adjusts the intratetrahedral interaction. 

For the comparison, we adjust both the complex $q$-deformed model ($q=e^{i\theta}$) and the undeformed model to the model of Eq.~\eqref{eq:H_tetra} with the experimental laboratory measurements of $J_1$ and $J_2$. {This demonstrates the validity of the deformed model, as it closely reproduces the experimental thermal behavior.}
The fitted parameters and the corresponding correlation coefficients are
summarized in Table~\ref{tab:qfits}. These have been chosen to obtain the better fit for all {thermal} observables for the same value of
$\theta$ and $I$. Fig.~\ref{fig:qfit} shows the {fit} for the Cu$_2$Te$_2$O$_5$Cl$_2$ compound, for both deformed and undeformed cases. 

\begin{table}[H]
\centering
\resizebox{\textwidth}{!}{
\begin{tabular}{l c c c c c | c c | c c | c c | c c}
\hline\hline
\textbf{Compound}
& $J_1$ [K]
& $J_2$ [K]
& $J_2/J_1$
& $g$
& $\theta$ [rad]
& $I(q)$ [K]
& $I(q=1)$ [K]
& $R^2_{C}(q)$
& $R^2_{C}(q=1)$
& $R^2_{\chi}(q)$
& $R^2_{\chi}(q=1)$
& $R^2_{M}(q)$
& $R^2_{M}(q=1)$
\\
\hline
Cu$_2$Te$_2$O$_5$Cl$_2$
& 37.0
& 45.0
& 1.216
& 2.03
& 0.422
& {-}46.56
& {-}45.42
& 0.9999
& 0.9575
& 0.9999
& 0.9963
& $\approx 1$
& 0.9995
\\
Cu$_2$Te$_2$O$_5$ClBr
& 26.0
& 37.0
& 1.423
& 2.03
& 0.537
& {-}39.50
& {-}37.84
& 0.9999
& 0.8702
& 0.9999
& 0.9928
& $\approx 1$
& 0.9974
\\
Cu$_2$Te$_2$O$_5$Br$_2$
& 31.5
& 46.0
& 1.460
& 2.07
& 0.553
& {-}49.40
& {-}47.34
& 0.9999
& 0.8818
& 0.9997
& 0.9867
& $\approx 1$
& 0.9966
\\
Cu$_2$Te$_2$O$_5$Cl$_{1.5}$Br$_{0.5}$
& 22.0
& 37.0
& 1.682
& 2.01
& 0.601
& {-}40.38
& {-}38.24
& 0.9999
& 0.7883
& 0.9995
& 0.9868
& $\approx 1$
& 0.9955
\\
\hline\hline
\end{tabular}
}
{\caption{Fitted parameters for the complex $q$-deformed model with $q=e^{i\theta}$ and the undeformed isotropic model $I(q=1)$ to the experimental data of the Cu$_2$Te$_2$O$_5$Cl$_2$ compound obtained in Ref.~\cite{jaglicic2006magnetic}. The correlation coefficients $R^2$ for the specific heat $C_{V}$, magnetic susceptibility $\chi$, and magnetisation $M$ compare the performance of the deformed and undeformed descriptions.}
\label{tab:qfits}}
\end{table}

\begin{figure}[H]
    \begin{minipage}{0.4\textwidth}
      \includegraphics[scale=0.9]{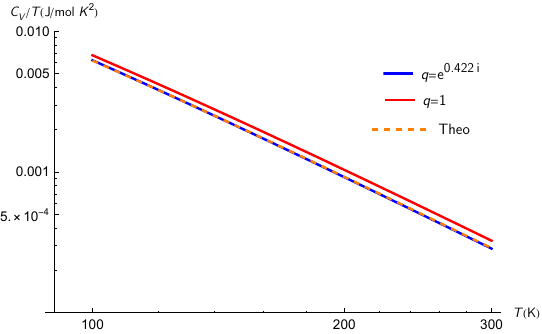}
    \end{minipage}
    \hspace{15mm}
    \begin{minipage}{0.4\textwidth}
      \includegraphics[scale=0.9]{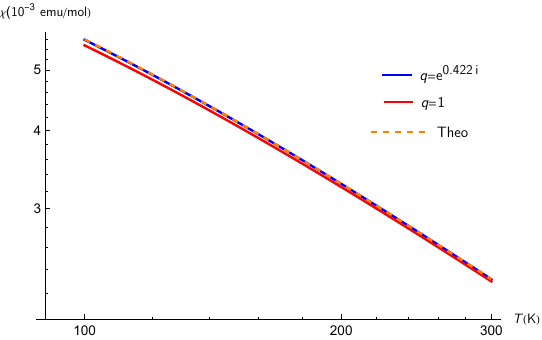}
    \end{minipage}
    {\caption{Specific heat divided by temperature, $C_{V}/T$ (left), and magnetic susceptibility, $\chi$ (right), as a function of temperature for the isolated tetrahedral cluster ($N=4$). The discontinuous lines correspond to the exact-diagonalization of the Hamiltonian \eqref{eq:H_tetra} for the Cu$_2$Te$_2$O$_5$Cl$_2$ compound, while red solid lines represent the complex $q$-deformed fit, and blue solid lines represent the $q=1$ case. 
    }
    \label{fig:qfit}}
  \end{figure}

The most relevant trend is the monotonic increase of $\theta$ with the
frustration ratio $J_2/J_1$. Indeed, the least distorted
case in the table, Cu$_2$Te$_2$O$_5$Cl$_2$, has the smallest value of
$\theta$, whereas the mixed compound Cu$_2$Te$_2$O$_5$Cl$_{1.5}$Br$_{0.5}$,
with the largest value of $J_2/J_1$, requires the largest deformation phase.

The improvement produced by the complex deformation is especially clear in
the specific heat. While the undeformed model already provides a reasonable
description of $\chi$ and $M$, it is much less accurate for $C_{V}/T$, where the
thermal population of low-lying levels is highly sensitive to small spectral
shifts. For example, in the mixed compound
Cu$_2$Te$_2$O$_5$Cl$_{1.5}$Br$_{0.5}$, the undeformed model gives
$R_C^2=0.7883$, whereas the complex $q$-deformed model gives
$R_C^2=0.9999$. This shows that the deformation mainly corrects the spectral
structure relevant to specific heat, magnetic susceptibility and magnetisation.

It is important to note that for smaller temperatures, the $q$-deformed model can also effectively describe the experimentally fitted model. For example, for the compound Cu$_2$Te$_2$O$_5$Cl$_2$, the {thermal} observables in the temperature range {$40 \leq T \leq 80~\mathrm{K}$} are accurately reproduced by the complex $q$-deformed model with {$\theta = 0.489$} and  {$I_q =- 46.92~\mathrm{K}$}. Even within this temperature window, the $q$-deformed description yields an excellent global agreement with the exact tetrahedral reference data. In particular, the specific heat is reproduced with a correlation coefficient {$R^2_C = 0.997$}, capturing both the position and the width of the broad maximum arising from short-range antiferromagnetic correlations inside the frustrated tetrahedron. The magnetic susceptibility also shows very good agreement ($R^2_\chi = 0.998$), correctly describing the smooth crossover from Curie-like behavior at high temperatures to the correlated regime at lower temperatures. For the magnetisation curve, we find $R^2_M \approx 1.0$. 

\begin{figure}[H]
    \begin{minipage}{0.4\textwidth}
      \includegraphics[scale=0.9]{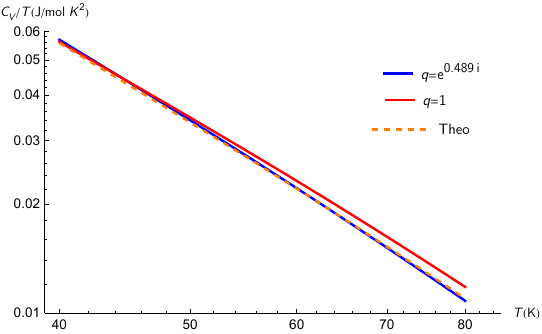}
    \end{minipage}
    \hspace{15mm}
    \begin{minipage}{0.4\textwidth}
      \includegraphics[scale=0.9]{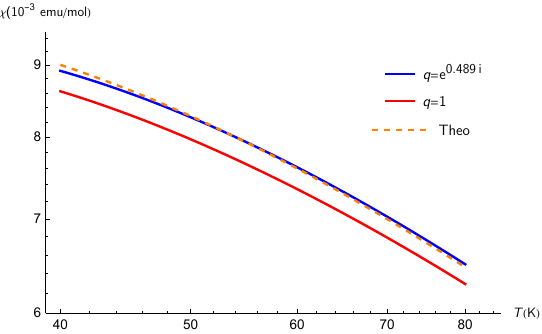}
    \end{minipage}
    {\caption{Specific heat divided by temperature, $C_{V}/T$ (left), and magnetic susceptibility, $\chi$ (right), as a function of temperature for the isolated tetrahedral cluster ($N=4$). The discontinuous lines correspond to the exact-diagonalization of the Hamiltonian \eqref{eq:H_tetra} for the Cu$_2$Te$_2$O$_5$Cl$_2$ compound, while red solid lines represent the complex $q$-deformed fit, and blue solid lines represent the $q=1$ case.}
    \label{fig:qfitlow}}
  \end{figure}

For comparison, the undeformed model leads to a smaller effective exchange scale {$I_{q=1} =- 45.50~\mathrm{K}$} and a visibly degraded description of the {thermal} quantities, particularly the susceptibility, where the correlation coefficient drops to {$R^2_\chi \simeq 0.892$}. Although the undeformed model remains continuous and qualitatively reasonable (with {$R^2_C = 0.980737$} and {$R^2_M = 0.999886$}), it fails to fully account for the redistribution of spectral weight produced by the geometrical frustration encoded in the ratio $J_2/J_1$.
}

{Finally, it is worth noting that for small clusters such as $N=4$ and $N=5$---where the undeformed model does not, in the latter case, correspond to an equi-spaced spin geometry---the presence of only two interaction length scales means that other two-parameter long-range models could capture this coupling scheme. This includes Inozemtsev's model~\cite{inozemtsev1990connection} or the $q$-deformation introduced by Lamers~\cite{lamers2022spin}. The distinctive features of the $q$-deformed Kittel--Shore model become more apparent from $N=6$ onwards, where the $U_q(\mathfrak{su}(2))$ symmetry constrains multiple interaction channels simultaneously. Nevertheless, the quantitative agreement obtained here confirms the physical viability of our approach for $\text{Cu}_2\text{Te}_2\text{O}_5X_2$, providing a natural starting point for future investigations into larger spin clusters.}

\section{Conclusions}
\label{sec:conclusions} 
In this work, the main thermodynamical properties of the $q$-deformation of the KS model were investigated. Introducing the quantum algebra $\mathfrak{su}_{q}(2)$, the main properties analysed in~\cite{al1998exact} for the KS model are studied here for its $q$-deformed version. In particular, for the $q$-deformed KS Hamiltonian, we studied the specific heat, magnetic susceptibility, magnetisation, phase transitions, and Curie temperature. Additionally, a systematic finite-size scaling analysis confirmed that deformation alters the critical temperatures and maximum amplitudes, but preserves the overall universality and scaling exponents.

In the ferromagnetic case, we cannot deduce approximate analytical expressions for the thermodynamic quantities. The difference with respect to the antiferromagnetic case, for which we were able to do it, is that the distance between the smallest energy levels tends to zero in the antiferromagnetic case. However, this distance is not negligible in the ferromagnetic case. This implies that the specific heat shows its maximum at a significantly higher temperature for larger $N$, in contrast to the antiferromagnetic case. Therefore, while the approximation of considering the two lowest energy levels reproduces the behaviour of the specific heat at low temperatures, it cannot describe the maximum, since it is at a higher temperature at which the approximation is not valid.

When considering the thermodynamic limit, the change in the coupling $I \to I/N$ causes the energy distribution to compress. This modification, combined with a deformation inversely proportional to the number of particles, ensures the {extensivity} of the model. In the undeformed case, the behaviour {shows a weak} dependence on the number of particles: in the specific heat, the peaks are similar in magnitude and occur at the same temperature; in susceptibility, the Curie transition occurs at nearly the same temperature; and in magnetisation, the curves are practically identical. {In contrast}, in the deformed case, the behaviour changes drastically. Therefore, careful consideration is required when selecting the values of the deformation parameter. In this study, particularly small values of $q$ were chosen, which were significantly smaller than those used in the antiferromagnetic case, to observe progressive changes. In all three thermodynamic functions, a shift in behaviour toward higher temperatures was observed when deformation was introduced.

In the antiferromagnetic case, by considering the most probable levels, we can find a reasonable and useful approximation to obtain analytical expressions for both undeformed and deformed scenarios, from which we can understand the behaviour of the thermodynamic quantities and phase transitions. {These expressions converge to be the same, up to a factor $1/N$, for both undeformed and deformed cases.} Moreover, we can find (exact) analytical expressions of the magnetic field at which phase transitions occur for both deformed and undeformed cases, which is a study not explored in~\cite{al1998exact}. The most appreciable change between the deformed and undeformed cases is observed in the magnetisation, particularly in the jumps at zero temperature as a function of the external magnetic field.

We also discussed the thermodynamic limit of the antiferromagnetic case. We observe that the analytical expressions of the thermodynamic quantities considered converge to be the same, up to a factor $1/N$, for both undeformed and deformed cases. This remarkable tendency disappears when deformation is introduced. The distribution of the energy states changes in the deformed case. Specifically, the excitation energies increase under deformation, so accessing the excited states requires more energy than in the undeformed case, which enhances the relative stability of the ground states. Consequently, the convergence may not disappear, but it would occur at temperatures higher than those considered, for which the susceptibility becomes very small (compared to its value at low temperatures).

{Furthermore, a systematic finite-size scaling analysis was carried out to characterize the critical behavior and elucidate how the system approaches the thermodynamic limit under different deformation regimes. In the ferromagnetic case, when a sufficiently large deformation is applied, it is found that both the specific heat and the magnetic susceptibility scale with $N$ at the critical temperature (corresponding to the maximum of the specific heat). Conversely, the finite-size response in the antiferromagnetic regime is strongly governed by system parity at low temperatures due to the fundamentally different structures of the even and odd ground-state sectors. Nevertheless, our scaling analysis demonstrates that both parity sectors asymptotically converge to a universal scaling law independently of the value of $q$, where the specific heat peak decays as $N^{-1}$ and the magnetic susceptibility approaches a size-independent limit, proving that the overall structure of the scaling laws remains robust against algebraic deformations.}

{Finally, we showed the changes in the eigenvalues of the deformed model and the {thermal} quantities for complex $q$,  not being $q$ a root of unity.  It is important to note that for complex  $q$ (not a root of unity), all the analytical results obtained in this work are  also valid.} 
{However, when $q$ is a root of unity, the underlying representation theory induces higher state degeneracies and distinct energy-level crossings, even though the explicit construction of the $q$-KS model remains unchanged~\cite{biedenharn1995quantum}.} 
{Thus, we confined our study to the case $N=4$ with $q$ on the unit circle, comparing this deformed model with another one commonly used in the literature to describe intratetrahedral interactions in the Cu$_2$Te$_2$O$_5$X$_2$ compound family. The numerical results suggest that the complex $q$-deformed framework effectively captures the thermodynamics of frustrated tetrahedral spin clusters, successfully describing the thermodynamics of the non-isotropy model. Consequently, the isolated microscopic tetrahedral Hamiltonian and the $q$-deformed effective model are not dynamically identical. Instead, they are thermodynamically equivalent within the temperature and magnetic field ranges considered in the fits, accurately reproducing macroscopic observables despite their differing local structures.}

{
It should be emphasized that this equivalence applies to very specific situations. Our analysis focuses strictly on isolated tetrahedral units. A full quantitative description of real materials would require the explicit inclusion of intertetrahedral couplings and lattice contributions.
Within these particular constraints, the complex $q$-deformed model serves as an effective tool to reproduce the {thermal} behavior of certain specific frustrated quantum magnets. This suggests that quantum-group structures can be useful for describing particular frustrated magnetic systems under some restrictive conditions.}

{As a final remark, while this study demonstrates the successful application of the $q$-deformed Kittel--Shore model to $N=4$ clusters in the family of $\text{Cu}_2\text{Te}_2\text{O}_5X_2$ compounds, alternative models could also describe these small clusters. The model's distinctive features will become more manifest for $N \ge 6$, where $U_q(\mathfrak{su}(2))$ symmetry imposes non-trivial constraints across multiple interaction channels simultaneously. The present results thus provide a solid foundation for future investigations into larger spin systems.}

\section*{Acknowledgments}
The authors appreciate the useful discussions with Ángel Ballesteros, Nicolás Cordero, and Iván Gutierrez-Sagredo. This work has been partially supported by Agencia Estatal de Investigaci\'on (Spain) under grant PID2019-106802GB-I00/AEI/10.13039/501100011033, by the Regional Government of Castilla y Le\'on (Junta de Castilla y Le\'on, Spain), and by the Spanish Ministry of Science and Innovation MICIN and the European Union NextGenerationEU (PRTR C17.I1). The authors would like to acknowledge the contribution of the COST Action CA23130. The authors have benefited from the activities of COST Action CA23115: Relativistic Quantum Information, funded by COST (European Cooperation in Science and Technology).

{
\appendix
\section{Level crossings and roots of unity}
\label{app:crossings}
The periodic nature of $q$-numbers with $q=e^{i\theta}$ induces exact energy level crossings. In a $q$-deformed KS model without an external magnetic field ($h=0$) and fundamental spin $j=1/2$, the energy of a multiplet with total spin $J$ follows from Eq.~\eqref{eq:qenergy} as:
\begin{equation}
E(J) = -\frac{I}{2}\left( \frac{\sin(J\theta/2) \sin((J+1)\theta/2)}{\sin^2(\theta/2)} - N \frac{\sin(\theta/4) \sin(3\theta/4)}{\sin^2(\theta/2)} \right).
\end{equation}
The exact values of $\theta$ where two distinct spin levels $J_1$ and $J_2$ ($J_1 \neq J_2$) share the same energy are found by setting $E(J_1) = E(J_2)$. Since the $N$-dependent term acts as a uniform shift for all energy levels, it cancels out exactly. Assuming $\theta$ is not a multiple of $2\pi$, and the identity
\begin{equation}
\sin(J\theta/2) \sin((J+1)\theta/2) = \frac{1}{2} \left[ \cos\left(-\frac{\theta}{2}\right) - \cos\left(\frac{2J+1}{2}\theta\right) \right],
\end{equation}
the equality reduces to:
\begin{equation}
\cos\left(\frac{2J_1+1}{2}\theta\right) = \cos\left(\frac{2J_2+1}{2}\theta\right).
\end{equation}
For two cosines to be equal, their arguments must satisfy $X = \pm Y + 2\pi k$ for any integer $k$. The negative branch generates the physical crossings within the energy spectrum:
\begin{equation}
\frac{2J_1+1}{2}\theta = -\frac{2J_2+1}{2}\theta + 2\pi k.
\end{equation}
Solving for $\theta$ yields the fundamental two-band crossing theorem for the model:
\begin{equation}
\theta_{cross} = \frac{2\pi k}{J_1 + J_2 + 1}.
\label{eq:crossing}
\end{equation}
Conversely, the positive branch corresponds to the difference of the arguments:
\begin{equation}
\frac{2J_1+1}{2}\theta = +\frac{2J_2+1}{2}\theta + 2\pi k,
\end{equation}
which simplifies to an alternative condition for the phase:
\begin{equation}
\theta = \frac{2\pi k}{|J_1 - J_2|}.
\label{eq:crossing_diff}
\end{equation}
This demonstrates that any two spin bands will always intersect whenever the deformation phase satisfies either of these two branches, strictly governed by the sum or the absolute difference of their total spins.

At the critical point $\theta = \pi$ (corresponding to $q = -1$), the denominator of the $J$-dependent term simplifies completely since $\sin^2(\pi/2) = 1$, reducing the general equation to:
\begin{equation}
E(J) = -\frac{I}{2} \sin\left(J\frac{\pi}{2}\right) \sin\left((J+1)\frac{\pi}{2}\right) + E_0(\pi),
\end{equation}
where 
\begin{equation}
    E_0(\pi) = \frac{I}{2} N \frac{\sin(\pi/4)\sin(3\pi/4)}{\sin^2(\pi/2)} = \frac{IN}{4}
\end{equation}
represents the uniform energy shift evaluated at the critical point.
When the system comprises an even number of spins, the total spin $J$ is restricted to integer values. Evaluating the $J$-dependent part at $\theta = \pi$, we encounter an absolute mathematical constraint: for any integer $J$, either $J$ is even, or $J+1$ is even.
\begin{itemize}
    \item If $J$ is an even integer ($J = 2m$ for some integer $m$), then the first sine term becomes $\sin(m\pi) = 0$.
    \item If $J$ is an odd integer, then $J+1$ is an even integer ($J+1 = 2m$), and the second sine term becomes $\sin(m\pi) = 0$.
\end{itemize}
Therefore, the product of these two sine functions is universally zero:
\begin{equation}
E(J) = \frac{IN}{4}, \quad \text{for } J = 0, 1, \dots, N/2.
\end{equation}
Thus, for any system with even $N$, setting $\theta = \pi$ triggers a simultaneous macroscopic multi-crossing. All eigenvectors collapse into a degenerate state at $E=IN/4$.

When $N$ is odd, the allowed spin values are half-integers. We can parameterize this as $J = m - 1/2$, where $m$ is a positive integer.
Substituting this into the energy equation at $\theta = \pi$:
\begin{align}
E(m) - E_0(\pi) &= -\frac{I}{2} \sin\left(\frac{(m - 1/2)\pi}{2}\right) \sin\left(\frac{(m + 1/2)\pi}{2}\right), \\
E(m) - E_0(\pi) &= -\frac{I}{2} \sin\left(\frac{m\pi}{2} - \frac{\pi}{4}\right) \sin\left(\frac{m\pi}{2} + \frac{\pi}{4}\right).
\end{align}
Using the trigonometric identity $\sin(A-B)\sin(A+B) = \sin^2(A) - \sin^2(B)$, and knowing that $\sin^2(\pi/4) = 1/2$:
\begin{equation}
E(m) - E_0(\pi) = -\frac{I}{2} \left[ \sin^2\left(\frac{m\pi}{2}\right) - \frac{1}{2} \right].
\end{equation}
This yields exactly two possible energy eigenvalues depending on the parity of the index $m$:
\begin{itemize}
    \item If $m$ is odd ($J = 1/2, 5/2, 9/2, \dots$): $\sin^2(m\pi/2) = 1 \implies E = \frac{IN}{4} - \frac{I}{4} = \frac{I}{4}(N-1)$.
    \item If $m$ is even ($J = 3/2, 7/2, 11/2, \dots$): $\sin^2(m\pi/2) = 0 \implies E = \frac{IN}{4} + \frac{I}{4} = \frac{I}{4}(N+1)$.
\end{itemize}
Then, an arbitrarily large system with odd $N$ does not collapse into a single point, but rather bifurcates into exactly two flat energy bands.

For the tetrahedral cluster of four spins, the total spin can take{s} the values $J \in \{0, 1, 2\}$. The exact crossing points between the different magnetic sectors across the parameter space can be found directly by applying the two-band crossing condition derived earlier in Eq.\eqref{eq:crossing}. Setting $E(J_1) = E(J_2)$ leads to two branches (positive~\eqref{eq:crossing_diff}  and negative~\eqref{eq:crossing}) of solutions for the phase $\theta$. Evaluating both branches for the tetrahedral cluster ($J \in \{0, 1, 2\}$) in the principal interval $\theta \in (0, 2\pi)$:
\begin{itemize}
    \item \textbf{Singlet ($J=0$) and Triplet ($J=1$):} 
    The negative branch yields $\theta = \frac{2\pi k}{2} = \pi k \implies \theta = \pi$.
    The positive branch yields $\theta = 2\pi k$ (no internal solutions).
    \item \textbf{Singlet ($J=0$) and Quintet ($J=2$):} 
    The negative branch yields $\theta = \frac{2\pi k}{3} \implies \theta = 2\pi/3, 4\pi/3$. 
    The positive branch yields $\theta = \frac{2\pi k}{2} = \pi k \implies \theta = \pi$.
    \item \textbf{Triplet ($J=1$) and Quintet ($J=2$):} 
    The negative branch yields $\theta = \frac{2\pi k}{4} = \frac{\pi k}{2} \implies \theta = \pi/2, \pi, 3\pi/2$. 
    The positive branch yields $\theta = 2\pi k$ (no internal solutions).
\end{itemize}
This direct application identifies all secondary, non-simultaneous crossings within the spectrum while explicitly confirming that $\theta=\pi$ is the unique solution shared by all pairs simultaneously, perfectly verifying the accidental multi-crossing degeneracy. However, it must be emphasized that the crossing points cannot be physically taken into account, as they correspond to values of $\theta$ that yield roots of unity for the deformation parameter $q$. As previously established, at these singular points the representation theory of the underlying quantum algebra changes, and the standard construction of the energy spectrum is no longer applicable.}

\bibliography{references}
\end{document}